\documentclass[twocolumn]{aastex6}
\usepackage{psfig}
\usepackage{natbib}
\usepackage{epsfig}
\usepackage{rotating}
\usepackage{amsmath}
\usepackage{graphicx,color}
\usepackage{subfigure}
\usepackage{multirow}
\usepackage{amssymb}
\usepackage{comment}
\usepackage{textcomp}
\usepackage{pslatex}
\usepackage{cleveref}
\usepackage{graphicx,color,layout}
\usepackage{comment}
\usepackage[flushleft]{threeparttable}
\usepackage{ulem}
\usepackage{syntonly}
\newcommand{\pa}{J1741$-$0840}
\newcommand{\pb}{J1840$-$0840}
\newcommand{\psra}{PSR J1741$-$0840}
\newcommand{\psrb}{PSR J1840$-$0840}

\newcommand{\br}{}

\begin{document}

\title[On the nulling, drifting, and their interactions in PSRs \pa\ and \pb]{On the nulling, drifting, and their interactions in PSRs \pa\ and \pb}

\author{Gajjar, Vishal\altaffilmark{1}}
\affil{Xinjiang Astronomical Observatory, CAS, 150 Science 1-Street, Urumqi, Xinjiang 830011, China\\
Key  Laboratory of Radio Astronomy, Chinese Academy of Sciences, Nanjing 210008, China \\
Space Science Laboratory, 7--Gauss way, University of California, Berkeley 94710, USA}
\author{J. P. Yuan\altaffilmark{2}, R. Yuen, Z. G. Wen, Z. Y. Liu, N. Wang}
\affil{Xinjiang Astronomical Observatory, CAS, 150 Science 1-Street, Urumqi, Xinjiang 830011, China\\
Key  Laboratory of Radio Astronomy, Chinese Academy of Sciences, Nanjing 210008, China}

\altaffiltext{1}{vishalg@berkeley.edu}
\altaffiltext{2}{yuanjp@xao.ac.cn}

\begin{abstract}
We report detailed investigation of nulling and drifting behavior of two pulsars PSRs \pa\ and \pb\ observed 
from the Giant Meterwave Radio Telescope at 625 MHz. 
\psra\ was found to show nulling fraction (NF) of around 30$\pm$5\% while \psrb\ was shown 
to have NF of around 50$\pm$6\%. We measured drifting behavior from 
different profile components in \psrb\ for the first time with the leading component 
showing drifting with 13.5$\pm$0.7 periods while the weak trailing component showed 
drifting of around 18$\pm$1 periods. Large nulling do hamper accuracy of these
quantities derived using standard Fourier techniques. A more accurate comparison was drawn from driftband slopes, measured after sub-pulse modeling. 
These measurements revealed interesting sporadic and irregular drifting behavior in both pulsars. We conclude that 
the previously reported different drifting periodicities in the trailing component 
of \psra\ is likely due to the spread in these driftband slopes. 
We also find that both components of \psrb\ show similar driftband slopes within the uncertainties. 
Unique nulling--drifting interaction is identified in \psrb\ where, in most occasions, 
the pulsar tends to start nulling after what appears to be an end of a driftband. 
Similarly, when the pulsar switches back to an emission phase, in most occasions it starts 
at the beginning of a new driftband in both components. Such behaviors have not been 
detected in any other pulsars to our knowledge. We also found that \psra\ 
seems to have no memory of its previous burst phase while \psrb\ clearly 
exhibits memory of its previous state even after longer nulls for both components.
We discuss possible explanations for these intriguing nulling--drifting 
interactions seen in both pulsars based on various pulsar nulling models.
\end{abstract}


\section{Introduction} 
\label{intro}
Single pulses from some radio pulsars show remarkable phenomena, such as nulling,  mode--changing, and sub--pulse drifting. 
Pulsar nulling has been shown to occur in around 100  pulsars to date \citep{rit76,ran86,big92a,viv95,wmj07,gjk12}, with the 
phenomenon exhibits as an abrupt switch off in the the pulsed emission for a certain duration (hereafter: {\itshape null state})\cite[]{bac70}. 
The amount of null pulses, also quantified by the Nulling Fraction (NF)\footnote{NF represent fraction of null pulses in the data.},  ranges from less 
than 1\% to more than 95\% (e.g. PSR J1717$-$4054; \citealt{wmj07} and PSR J1752+2359; \citealt{gjw14a}). The nulls were also found 
to last for different durations ranging from just one or two seconds (e.g. PSR B0809+74; \citealt{rit76}) to many hours or even years 
(e.g. PSR J1832+0029; \citealt{crc+12}). 

In many long-period pulsars, the consecutive sub--pulses in a sequence of
single pulses (folded at the basic pulsar period P$_1$) vary in phase or
longitude systematically, forming apparent driftbands. The separation between 
driftbands where sub--pulses reappear at similar longitude, also known as drifting periodicity, P$_3$,  
varies from several to a few tens of P$_1$. The drift rate is defined as $D~ = ~ P_2/P_3$, 
where P$_2$ is the spacing between sub--pulses within
a single pulse. Some pulsars are known to switch between different P$_3$ modes. 
For example PSRs B0031$-$07 \cite[]{htt70} and B2319$+$60 \cite[]{wf81} 
exhibit three distinct values in the drift rates. 

In a few pulsars, both nulling and drifting phenomena were seen to co--exist and even shown to be 
interacting with each other. For example, PSRs B0809+74 and B0818$-$13 exhibit changes in the drift rate 
after the null states \cite[]{la83,vsrr03,jv04}. During some of these interactions, 
the pulsars also seem to indicate some sort of phase memory 
in that information was retained regarding the phase of the last burst or active pulse during the null state. 
For example, PSR B0031$-$07 was shown to retain memory of its pre-null burst  
phase across short nulls \cite[]{vj97,jv00}. If both 
nulls and drift--mode changes are result of intrinsic changes in the pulsar 
magnetosphere, then their interactions could provide important clue for understanding 
the triggering and transition mechanism for changes between different magnetospheric states. 
Furthermore, evidences have been mounting to support a close
relationship between nulling and drift--modes/profile--modes switching \cite[]{wmj07,lhk+10,cor13,hhk+13,gjw14a}. 
It is likely that nulling in itself could be an extreme form of mode--changing where
a pulsar switches between different magnetospheric states as indicated by recently 
reported broadband behavior of three nulling pulsars \cite[]{gjk+14b}. 
Thus, studies of pulsars that exhibit both nulling and drifting phenomena 
are essential in order to understand the true nature and origin of the nulling phenomenon.  
We present here observations of two pulsars, PSRs \pa\ and \pb, which 
appear to show interactions between the nulling and drifting phenomena as hinted from 
the single pulses reported by \cite{bjb+12}. However, these observations were not 
long enough to carry out detailed investigations of nulling and its interaction with 
the drifting phenomenon. Here, we report results for these pulsars from relatively longer observations, whose basic parameters are listed in Table \ref{intro_table}.

\begin{table}
\centering
 \caption{Basic parameters of two pulsars}
 \begin{threeparttable}
   
 \begin{tabular}{l c c}
 \hline
 PSR 			&	\pa\		&	\pb\	\\
 \hline
 \hline
 Period (sec)		&	2.043082	&	5.309377	\\
 DM (pc$\cdot$cm$^{-3}$ )	&	74.90 		&	272.00		\\
 RA (J2000)		&	17h41m22.5s	&	18h40m51.9s	\\
 DEC (J2000)		&	$-$08$^\circ$40'31.9''&	$-$08$^\circ$40'29'' \\
 Age (Myr)		& 	14.2		&	3.5	\\
 B$_s$ ($\times$10$^{12}$ G)&	2.18		&	11.3	\\
 $\dot{E}$ ($\times$10$^{30}$ ergs/sec)& 10.5	&	6.25	\\ 
 \hline 
 \end{tabular}
 \label{intro_table}
 \begin{tablenotes}[para,flushleft]
 \small
 \item Note : Here, the DM represents dispersion measure while the RA and DEC are the 
 right ascension and declination, respectively. The B$_s$ represents average surface magnetic field while the $\dot{E}$ represents 
 rate of loss--of--energy of the pulsar.
\end{tablenotes}
 \end{threeparttable}
\end{table}

\psra\ (PSR B1738$-$08) was discovered during the earlier searches \cite[]{mlt+78} from the Molonglo telescope
in Australia. It is a known drifting pulsar with a few studies \cite[]{wes06,wse07,bmm16} 
reporting its abnormal drifting behavior. It displays two distinct profile components at 
325 MHz. \cite{wes06,wse07} have reported drifting phenomenon 
in this pulsar for the first time at 1420 MHz and 325 MHz, respectively.  
\cite{wse07} hinted towards a possible nulling phenomenon in \psra, however, 
the length of their observation was not long enough to quantify the occurrence of nulls. 
To investigate nulling, drifting and their interaction further, we observed this pulsar for nearly two hours and obtained 
around 3095 single pulses. \cite{wes06} have reported different 
P$_3$ values for the leading and the trailing components in this pulsar at 1400 MHz. 
The trailing component was reported to have P$_3$ of around 4.4$\pm$0.3 periods while the leading component has 
slightly slower drifting with P$_3$ of around 5.2$\pm$0.1 periods. 
The authors suggested that this difference in P$_3$ 
in the two components could be due to drift--mode switching. \cite{wes06} have indicated 
that the trailing component might also have an additional drifting feature 
with P$_3$ of around 2 periods. Further observations at 300 MHz by \cite{wse07} revealed two P$_3$ values of 4.8$\pm$0.1 and 4.7$\pm$0.1 periods which were not 
possible to separate between the two components due to the sensitivity of their observations. 
Moreover, \cite{bmm16} have confirmed the 4.7$\pm$0.6 periods feature at 325 MHz 
and the 4.8$\pm$0.6 periods feature at 610 MHz. Here, we report its detailed behavior 
at 625 MHz with longer observations in contrast to \cite{bmm16}. 

\psrb\ was discovered during the recent searches conducted 
from the Parkes telescope \cite[]{lfl+06}. \cite{bjb+12} reported 
short single pulse observation of this pulsar which pointed 
toward possible occurrence of nulling and drifting phenomena. 
The period of this pulsar is very large ($\sim$5.3 sec) giving around 1260 single pulses in 
our 2-hour observation. This pulsar exhibits two profile components at 610 MHz, which are located very close 
to each other. The leading component is slightly stronger in intensity than that of the trailing component. 
We are reporting here a unique nulling-drifting interaction in this pulsar for the first time. 

The observations are discussed in Section \ref{observations} along with the 
descriptions on data recording and off-line data processing. The nulling behavior of both pulsars 
is discussed in Section \ref{nulling_sect} in which we discuss quasi-periodic feature 
found in \psra. Single pulse drifting results are discussed in Section \ref{drifting_sect}. 
A modeling of driftbands is discussed in Section \ref{drift_rate_sect} which is
essential to determine changes in the drifting pattern. We discuss null associated 
changes in these drift rates in Section \ref{null_induce_D_sect}, while the phase memory 
across the nulls is discussed in Section \ref{drift_phase_memory_sect} for both pulsars. 
The final conclusion and implication of our findings are discussed in the Section \ref{discussion_and_conclusion_sect}. 

\section{Observations}
\label{observations}
\begin{figure}
 \centering
 \subfigure[]{
  \includegraphics[width=8 cm,height=3 cm,angle=-90,bb=0 0 509 193,scale=1]{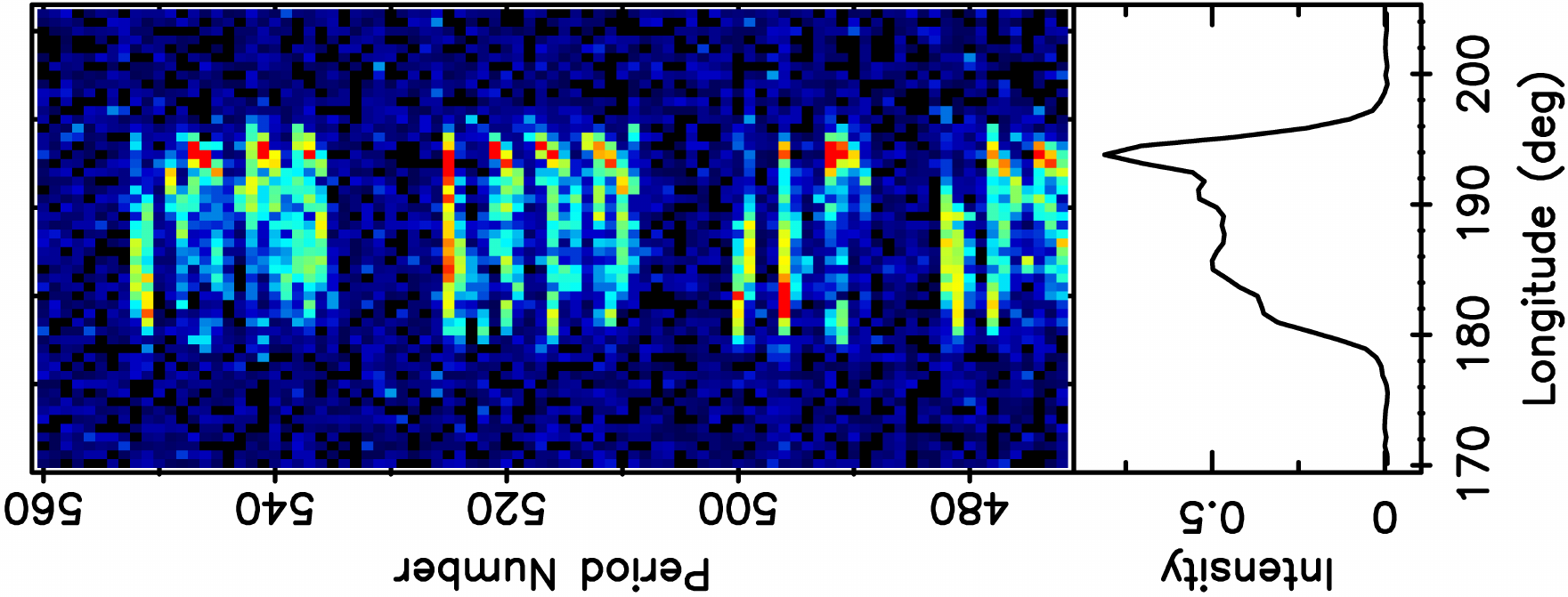}
  \label{1741_spdisplay}
  }
  \hspace{1cm}
  \subfigure[]{
  \includegraphics[width=8 cm,height=3 cm,angle=-90,bb=-10 0 499 193,scale=1]{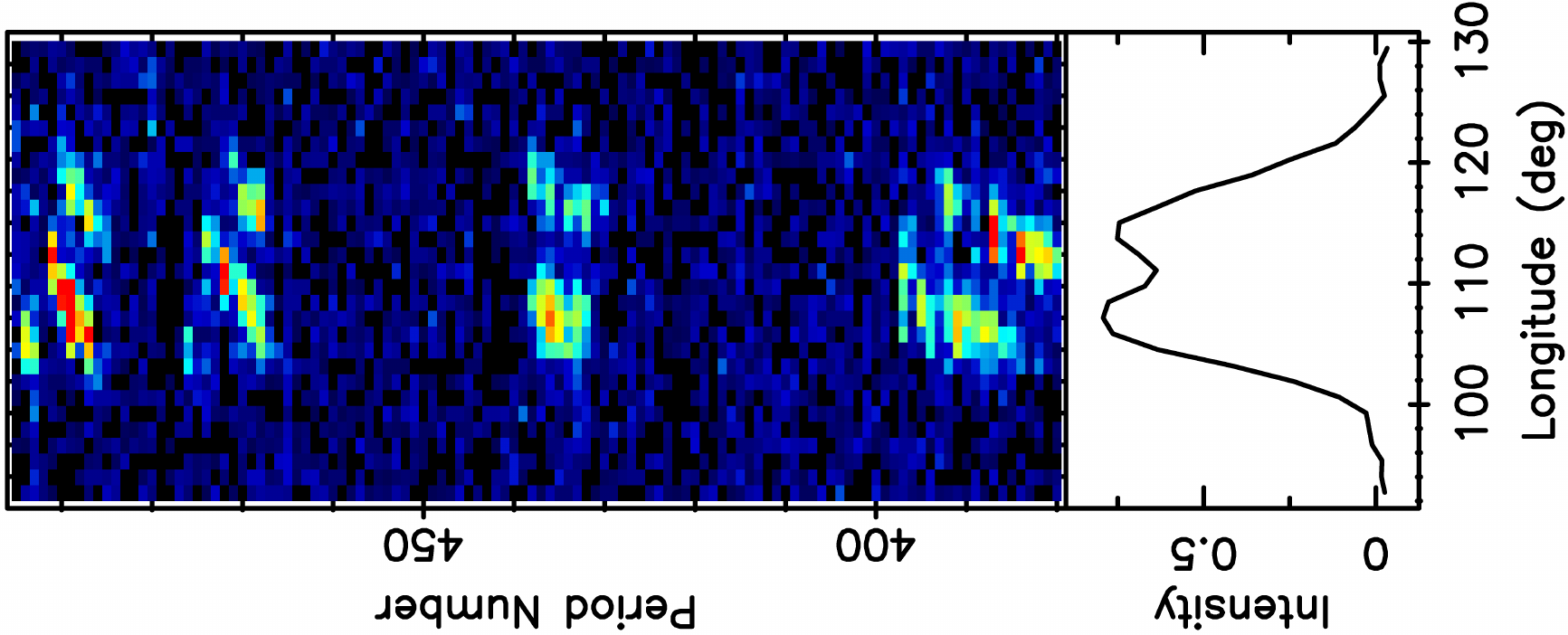}
  \label{1840_spdisplay}
  }
  \put(-190,0){\psra}
  \put(-65,0){\psrb}
  \caption{Sections of single pulses for the two pulsars observed at 625 MHz from the GMRT. The top panel  
  shows a section of observed pulses while the bottom panel shows the integrated profile from the entire observation for each pulsar. 
  (a) \psra\ shows prominent nulls separated by burst bunches which exhibit clear sub--pulse drifting in the trailing component, 
  (b) \psrb\ exhibit prominent long and short nulls along with 
   prominent drifting during the short burst phases.
  }
  \label{spdisplay_all}
\end{figure}

\begin{figure*}
 \centering
  \subfigure[]{ 
   \includegraphics[angle=-90,bb=0 0 504 720,scale=0.3]{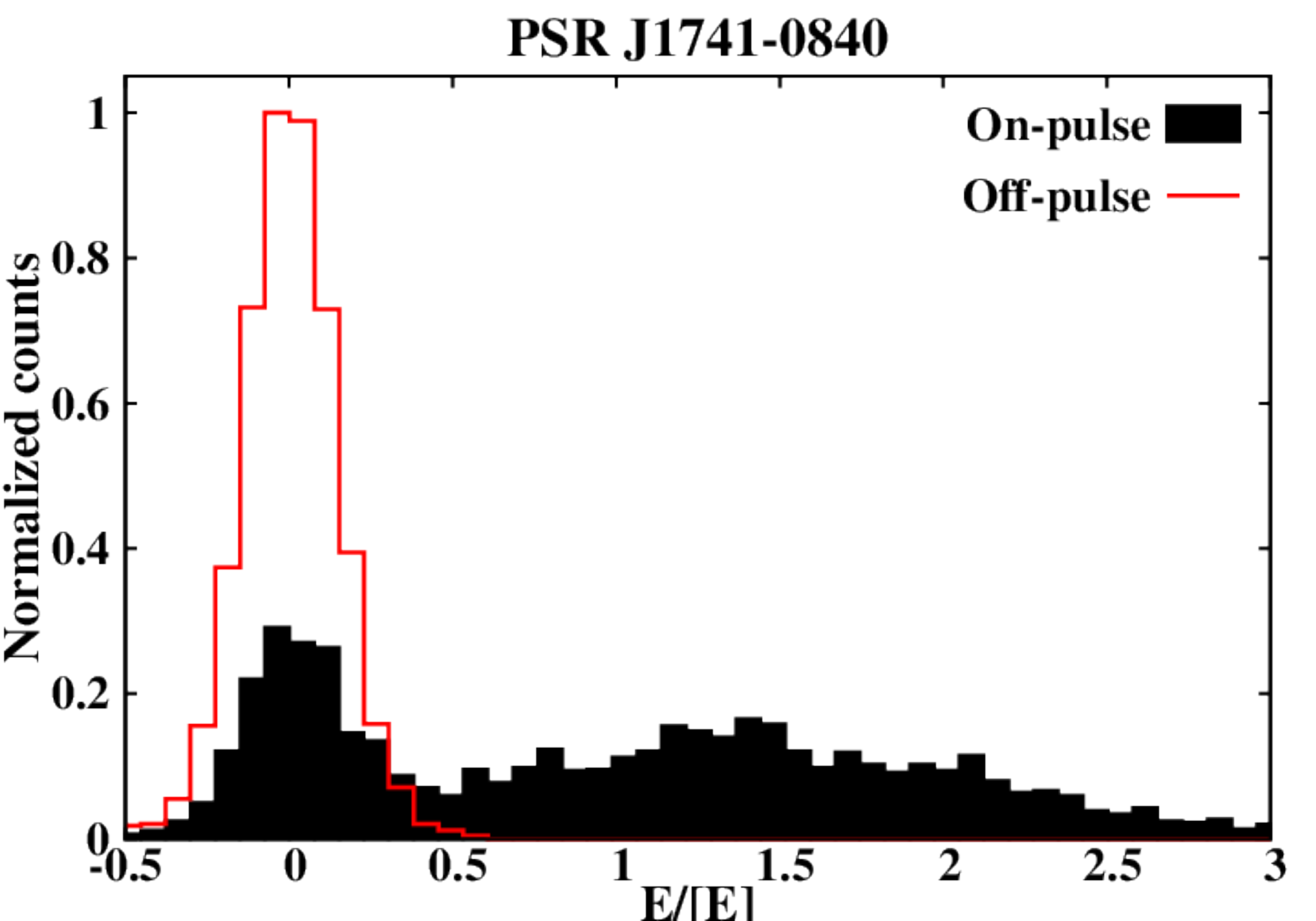}
   \label{1741_NF_hist}
  }
  \subfigure[]{ 
   \includegraphics[angle=-90,bb=0 0 504 720,scale=0.3]{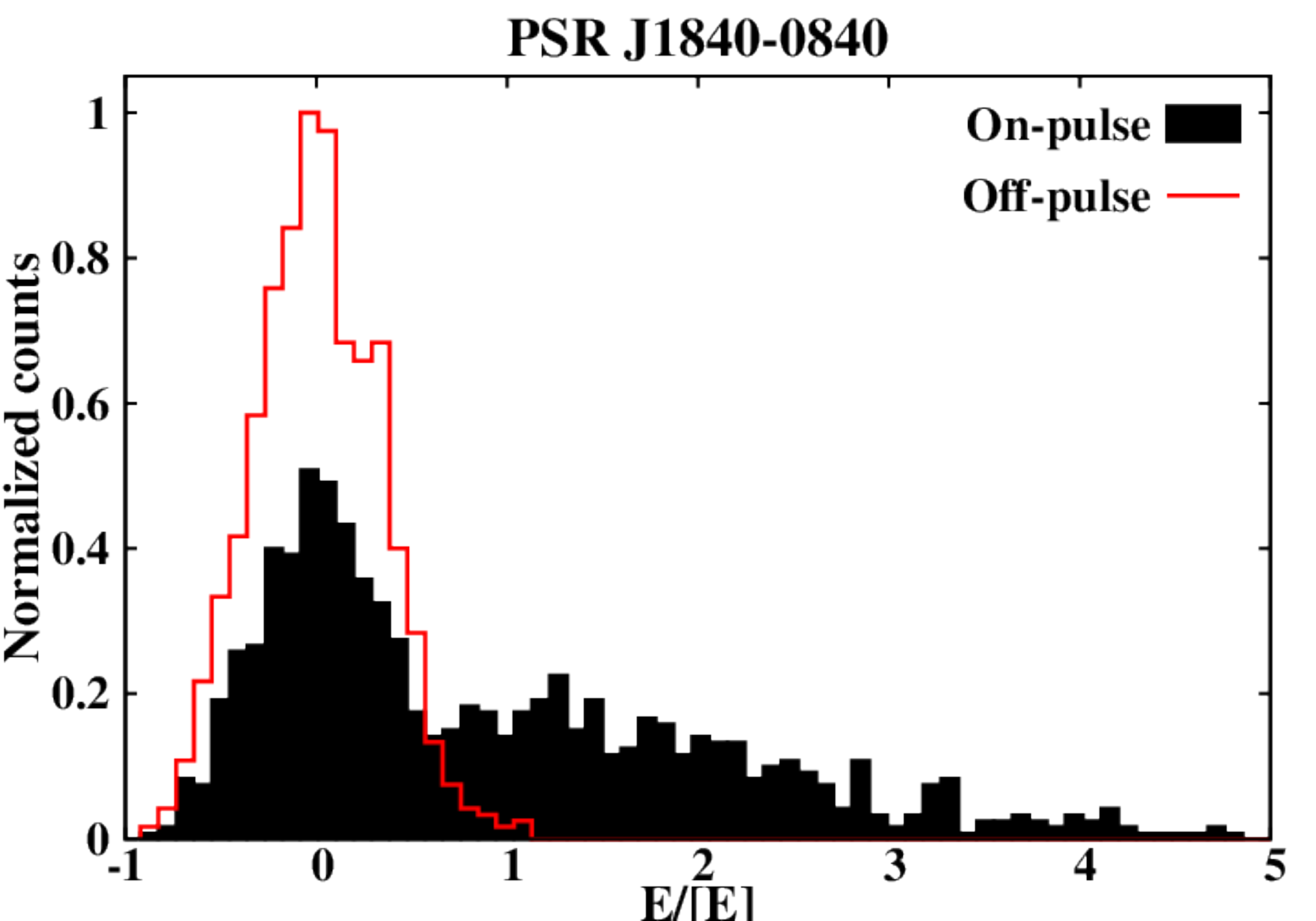}
   \label{1840_NF_hist}
  }
  \caption{The on-pulse and off-pulse energy histograms for the two pulsars observed from the GMRT at 625 MHz. 
 The red solid line indicates off-pulse histograms while the black fill regions are the on-pulse energy 
 histograms. (a) PSR J1741$-$0840 shows very clear nulling with a distinctly separate null and burst pulses 
 among the on-pulse energy distribution. The estimated NF for this pulsar is around 30$\pm$5\%. (b) \psrb\ 
 exhibits large fraction of null pulses, with NF of around 50$\pm$6\%, which can also be noticed from the large 
 distribution of pulses around zero pulse energy.}
 \label{NF_hist}
\end{figure*}
The observations were conducted using the Giant Meterwave Radio Telescope \cite[]{sak+91} 
on 19th January, 2013. The GMRT correlator was used as a digital filterbank to obtain 
512 spectral channels, each having a bandwidth of 62.5 KHz across the 
32 MHz bandpass received for each polarization from each GMRT antenna. 
The digitized signals from the 14 GMRT antennas were added in phase 
to form a coherent sum of signals with the GMRT Array Combiner (GAC) \cite[]{pr97}.
Then, the summed signals were detected in each channel of each polarization.  
The detected power in each channel were then acquired into 16-bit register every 
122.88$\mu$s after summing the two polarizations in a digital backend
and were accumulated before being written to an output buffer to reduce
the data volume. The data were then acquired using a data acquisition 
card and recorded to a tape$-$drive for the off-line processing. 

In the off-line processing, the raw filterbank data were first processed to handle 
spurious radio frequency interferences (RFI). We have deployed a similar method as suggested 
by \cite{hvm04} and \cite{vdo12} to remove time-domain and frequency-domain 
RFI from the raw GMRT filterbank data. We first determined median from 512 frequency channels 
for each time sample. Channels which exhibit three times more power than 
the median value were identified and replaced with the measured median value for 
a given time sample. For the time-domain RFI excision, we first calculated number of samples 
corresponding to the period of the pulsar in a given file. We calculated average power 
by adding all frequency channels for each sample. Samples that exhibited powers larger than 
three times the median value were replaced with a combination of a median and Gaussian noise 
equivalent to the standard deviation of the data. In order to keep the 
sensitivity similar for each observed pulse, similar number of samples were replaced 
for each period. For each block $-$ of number of samples per period and 
number of frequency channel per sample $-$ only top 20\% of 
the {\itshape bad} data were replaced. The raw GMRT filterbank data were then converted to 
SIGPROC\footnote{http://sigproc.sourceforge.net}\cite[]{lor01} filterbank format. 
We used standard SIGPROC dedispersion and folding tool to produce single pulses 
for further analysis of each pulsar. The data were folded to 512 or 1024 phase bins across the topocentric pulse 
period to obtain the single pulse sequences. It should be noted that although 
excision of RFI from the raw filterbank data improved the S/N for the observed single pulses, there were a 
few pulses which showed clear presence of RFI and hence removed entirely for the nulling fraction (NF) analysis of both the pulsars.

\section{Nulling}
\label{nulling_sect}
\begin{figure}
 \centering
 \subfigure[]{
  
  \includegraphics[width=8 cm,height=5 cm,angle=0,bb=0 0 842 595]{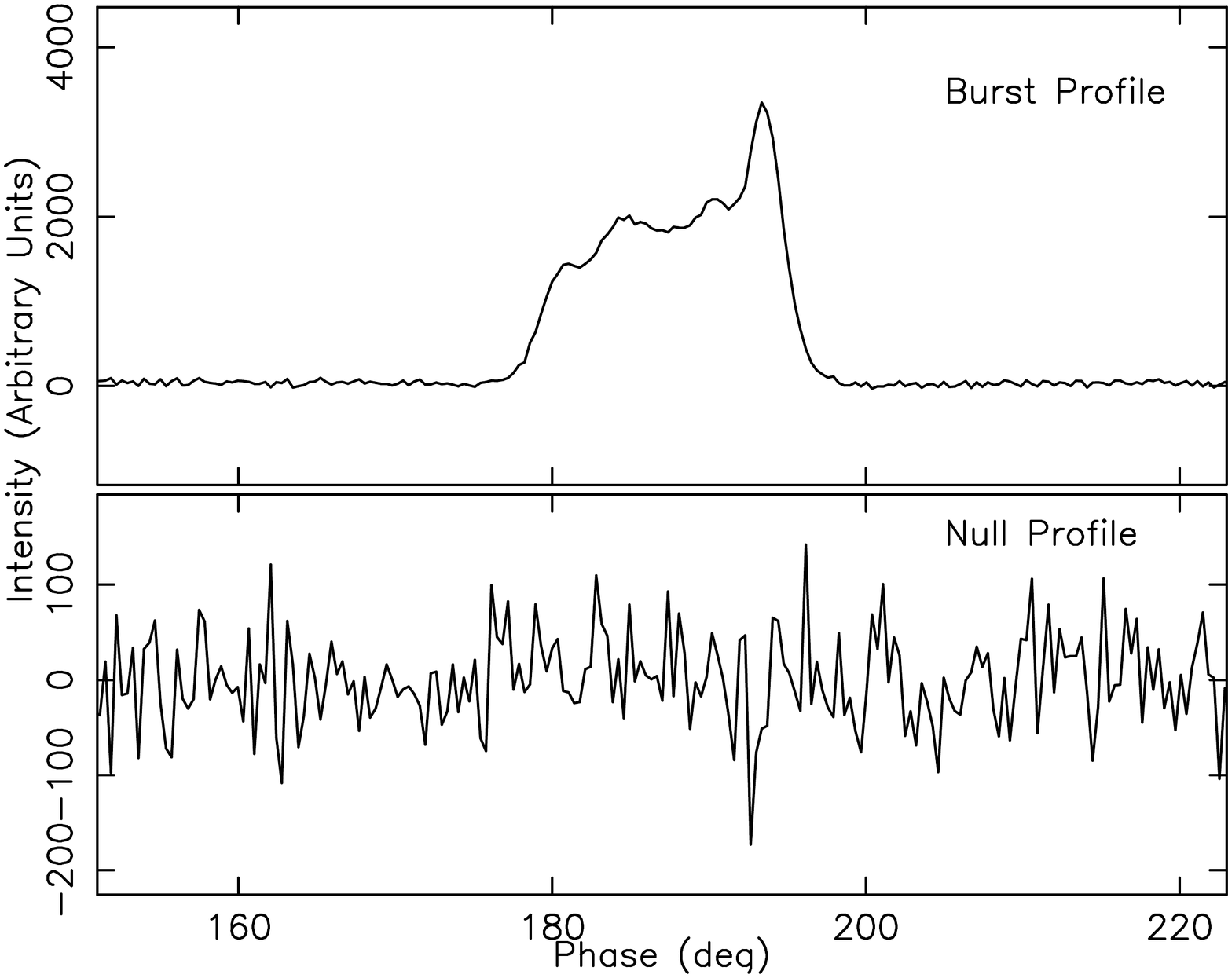}
  \label{NLBL_prof_1741}
  }
  \subfigure[]{
   \includegraphics[width=5 cm,height=7 cm,angle=-90,bb=0 0 560 707]{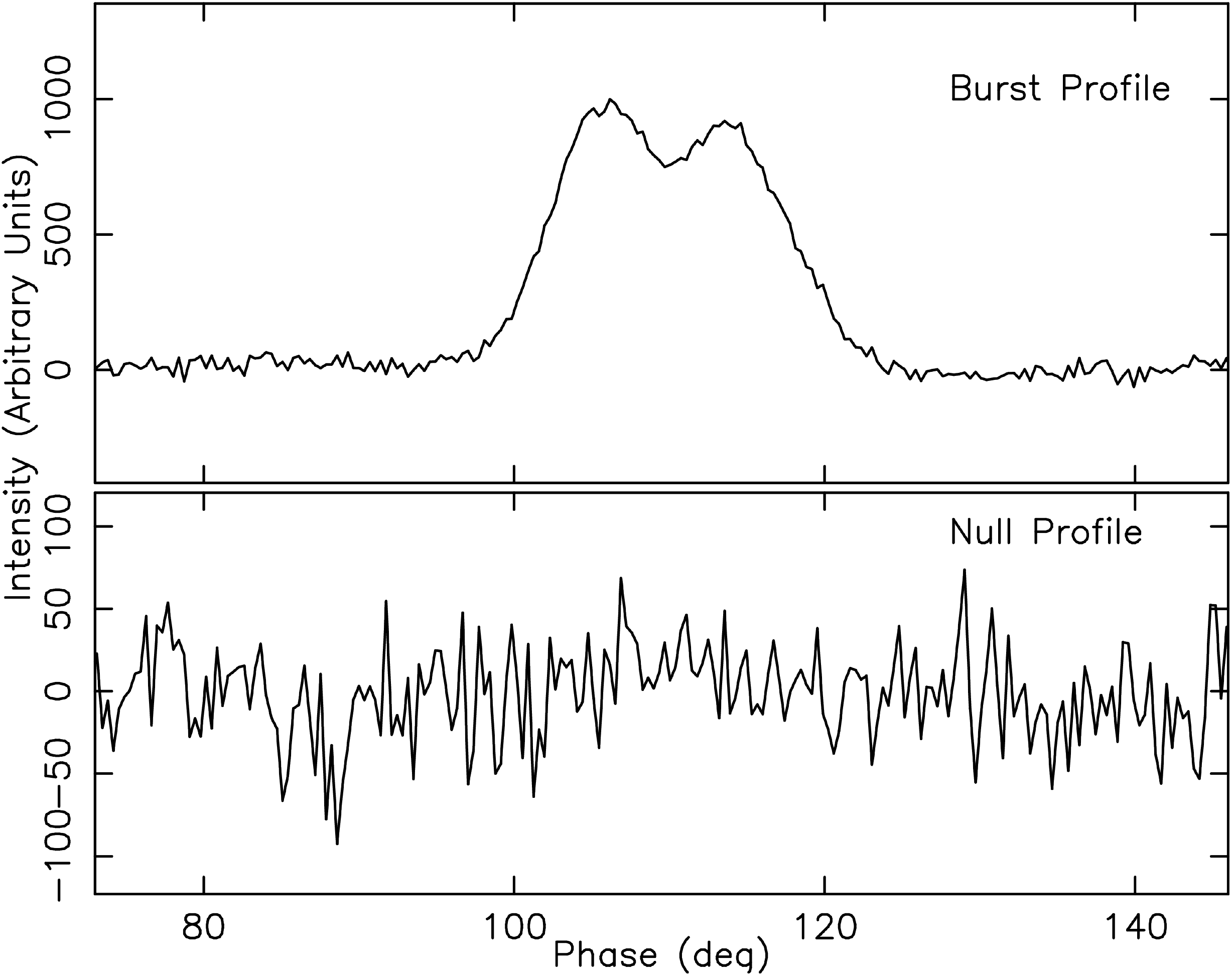}
  \label{NLBL_prof_1840}
  } 
  \caption{The integrated profile obtained separately for null and burst pulses for (a) \psra\ and  
  (b) \psrb. The top panel shows profile from the burst pulses only and the bottom panel shows 
  average profile after combining only null pulses for respective pulsars. The null profile 
  does not show any detectable profile component, suggesting absence of any weak level emission.}
 \label{NLBL_prof_both}
\end{figure}

\begin{figure}
 \centering
 \subfigure[]{
  \includegraphics[width=8 cm,height=6 cm,angle=0,bb=0 0 842 595,scale=1]{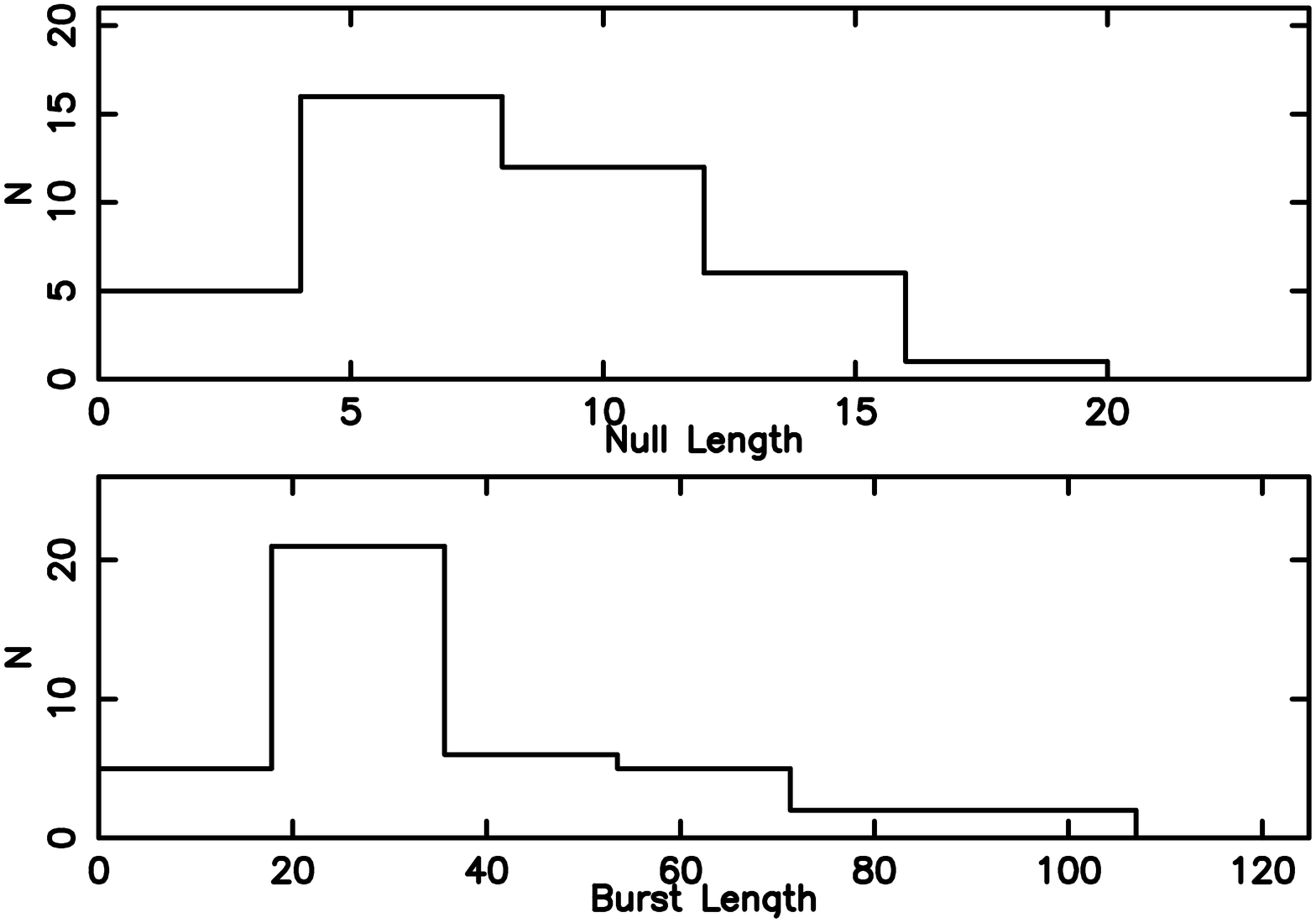}
  \label{NL_BL_1741}
  }
 \subfigure[]{
  \includegraphics[width=5 cm,height=7 cm,angle=-90,bb=0 0 566 760]{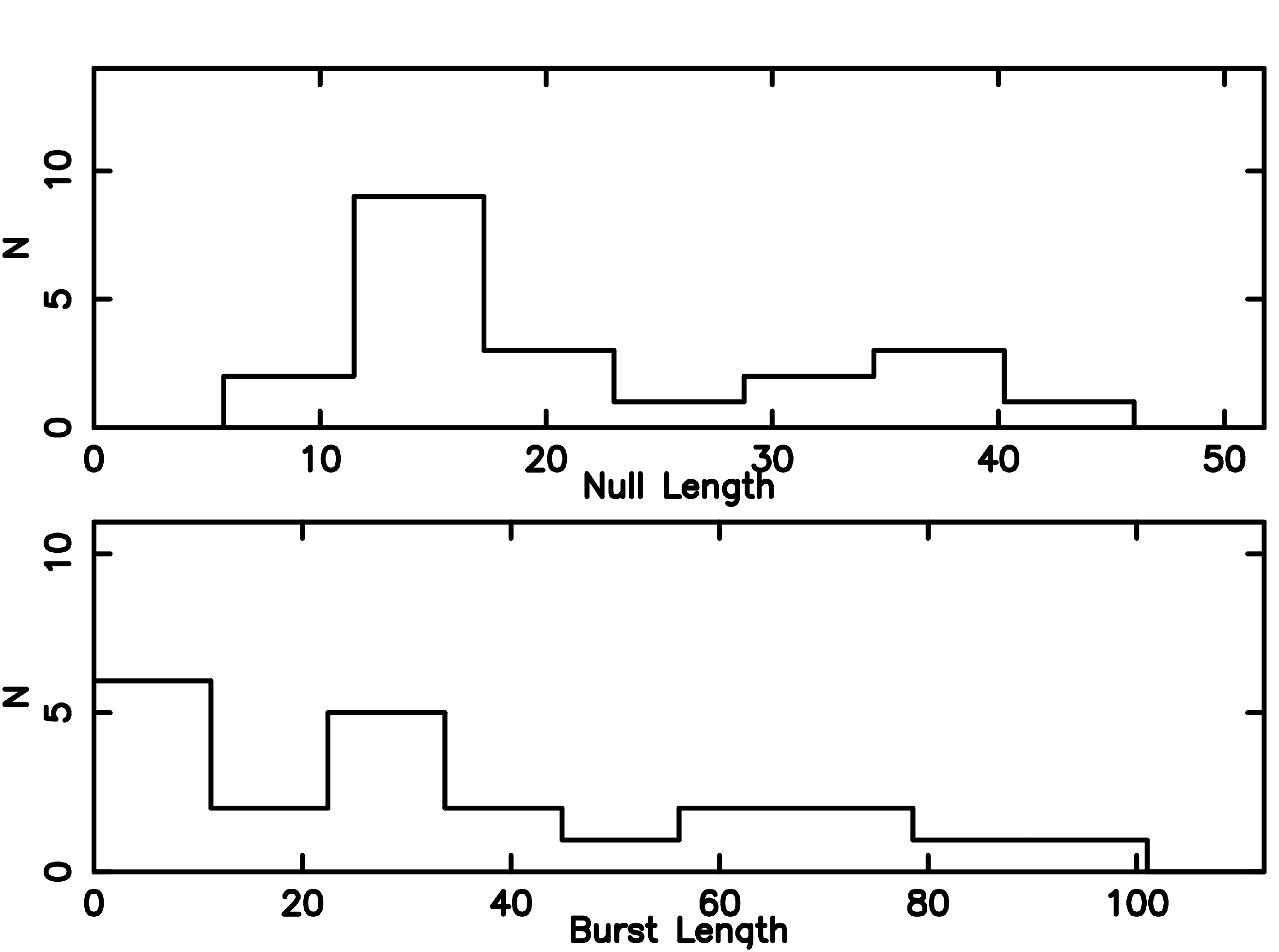}
   \label{NL_BL_1840}
 }
 \caption{The null$-$length and burst$-$length distributions for (a) \psra\ and (b) \psrb. The top panel shows distribution of 
 length of observed nulls while the bottom panel shows distribution of observed consecutive burst pulses for corresponding pulsars.
 \psra\ shows excess of around 30 period bursts corresponding to quasi-periodic feature seen from the burst bunches. For \psrb\ an excess 
 can be seen around null lengths of 13 to 15 periods.}
\end{figure}

We report here the NF for both pulsars for the first time. 
Figure \ref{spdisplay_all} shows a section of the observed single pulses, in which the pulses with no detectable 
emission are clearly visible from the two pulsars. It should be noted from Figure \ref{1741_spdisplay} 
that $-$ unlike profile at 325 MHz \cite[]{wse07} $-$ \psra\ does not exhibit two distinct profile components 
at 625 MHz. Prominent nulling along with regular drifting in two components can be seen for \psrb.
Figure \ref{1840_spdisplay} also shows an example of occasional long null of around 27 periods in \psrb.  
The NFs for both pulsars were obtained using a procedure similar to that 
used for detecting pulse nulling from the single pulse sequences by \cite{gjk12}. 
We obtained the on--pulse and off--pulse energy histograms using appropriate 
on--pulse and off--pulse windows of similar length. Figure \ref{NF_hist} 
shows these histograms with prominent nulls located near the zero pulsed energy for both pulsars. 
In Figure \ref{NF_hist} the on--pulse energy histograms show clear bi--modal distributions that 
point toward two separate populations each belonging to null and burst pulses for both pulsars. 
A Gaussian function was fitted for both the on-pulse and off-pulse histograms for each pulsar. 
A ratio between the peaks of on--pulse and off--pulse histograms provides estimation of 
NF\footnote{Although \cite{gjk12} have demonstrated that the NF is not an ideal parameter to quantify nulling, 
we are reporting them here for the sake of comparison.} for a given pulsar. The estimated NF for \psra\ is around 30$\pm$5\% 
while the estimated NF (50$\pm$6\%) for \psrb\ suggests that around half the observed pulses 
are likely to be nulls. The larger error bars are due to weak burst pulses which do hinder a clear distinction 
between these populations.

It is essential to confirm  that the observed nulls are true nulls and are not just weak 
emission phases where emission from the single pulses goes below the detection threshold of the telescope. 
For example, PSRs B0826$-$34 and B0823+26 were reported to show weak emission in the 
previously reported nulls \cite[]{elg+05,syw15}. We separated all the nulls from the burst pulses using a method described in 
\cite{gjk12,gjw14a} for both pulsars. A weighted sum was performed over all on--pulse bins for each observed pulses using 
the shape of their integrated pulse profile. For each pulsar, all the observed pulses were arranged in the ascending order of 
their weighted pulse energy. A threshold was moved from the high energy end towards the low energy end until pulses below this threshold 
did not show a significant profile component ($<$3$\sigma$). This threshold was used to divide 
all pulses into two groups with pulses below the threshold marked as nulls and pulses above the threshold marked as bursts. 
A visual inspection was also carried out on all the separated null pulses and most of the mislabeled burst pulses were removed and added to 
the burst pulse group. Figure \ref{NLBL_prof_both} confirms that these nulls clearly show absence of 
any detectable emission for all components in both pulsars. 


\subsection{Quasi-periodic fluctuations}
\label{qperiodicity_section}
\label{NLBL_section}
The separated null and burst pulses were again arranged in the order they were observed for each pulsar. 
It should be noted that as both pulsars exhibit multiple profile components, we consider complete 
absence of emission for all components as the beginning of a null state. 
The length of contiguous emission states (in any profile component) were measured and 
labeled as {\itshape burst$-$lengths} and separation between these bursts were measured as {\itshape null$-$lengths}. 
For \psra, as shown in Figure \ref{NL_BL_1741}, the burst--lengths clearly shows a good fraction of 
burst bunches with 25 to 30 periods while the null--lengths tends to show cluster between 5 to 10 periods.
\psrb\ shows clustering at null length of around 13 to 15 periods while the burst length seems to have
short bursts along with bunches around 25 to 30 periods. 
In earlier studies, a few nulling pulsars were known to show such clustering \cite[]{rw08,gjw14a}, 
which could give rise to quasi-periodic pulse energy fluctuations. 

\begin{figure}
 \centering
 \subfigure[]{ 
 
 \includegraphics[bb=0 0 360 252,scale=0.54]{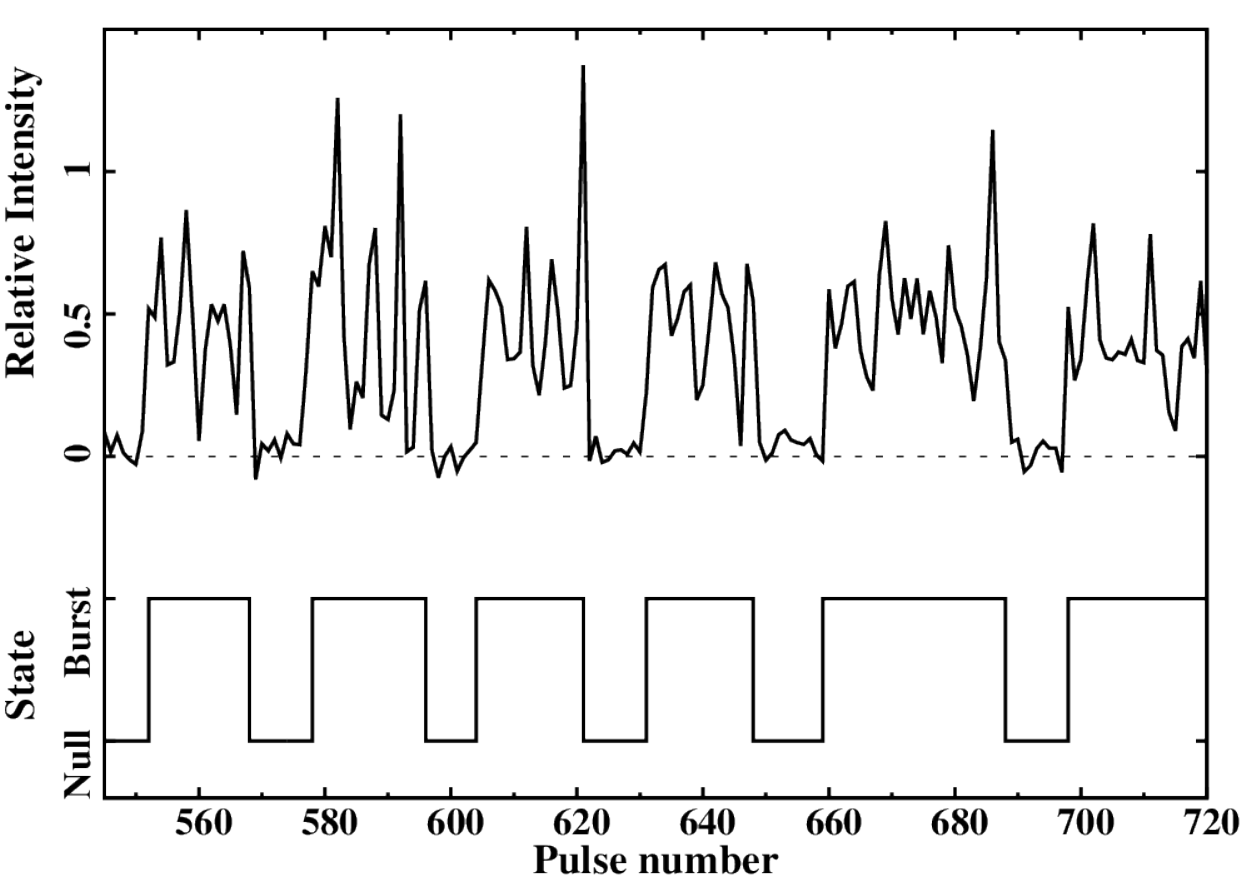}
 \label{1741_onps}
 }
 \subfigure[]{ 
 \includegraphics[bb=0 0 360 252,scale=0.54]{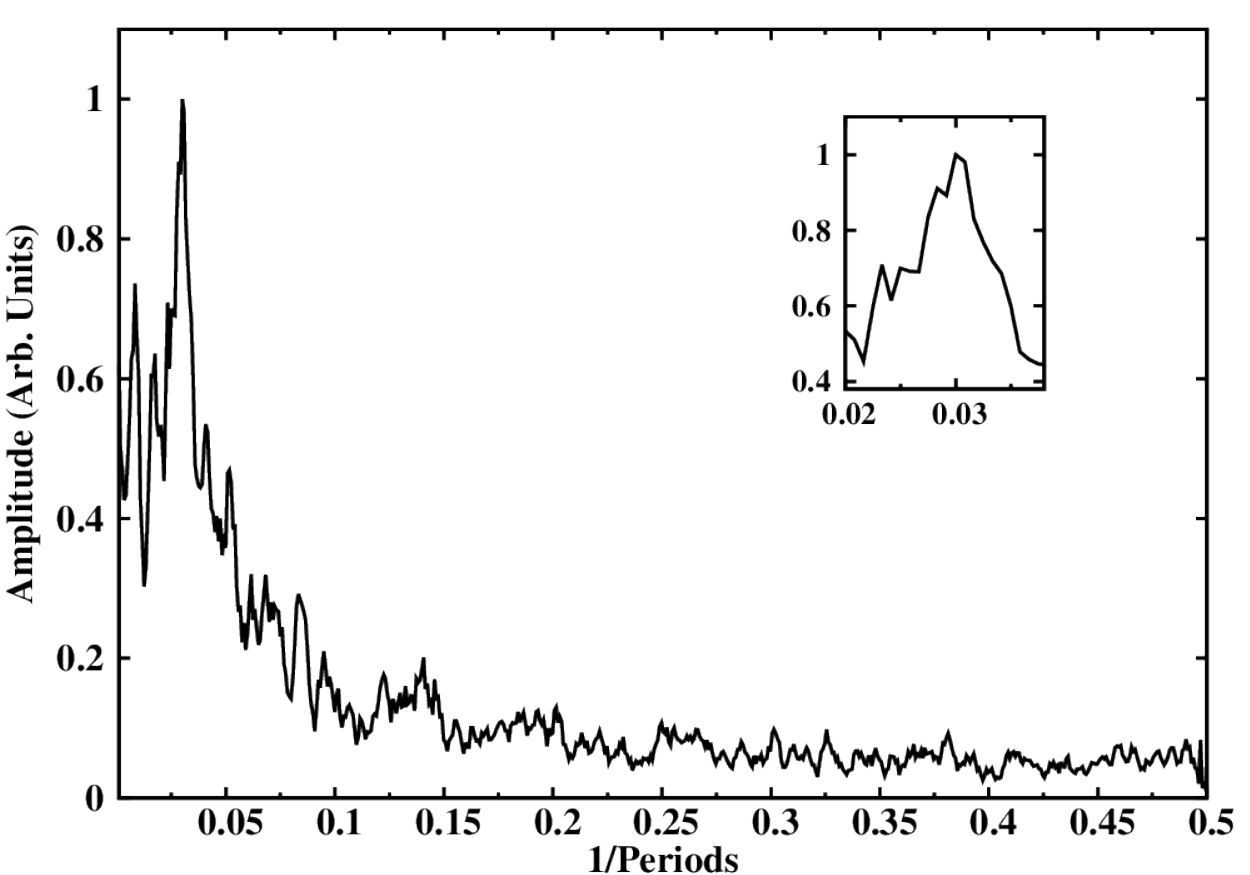}
 \label{1741_fft}
 }
 \caption{Quasi-periodic fluctuations in the pulse energy of \psra  at 625 MHz. (a) The top panel shows 
 the pulse energy modulation as a function of observed period. The bottom panel shows the 
 identified emission state (null or burst) corresponding to each period. (b) Fourier spectrum 
 of the pulse energy modulation pattern which clearly indicates strong periodicity. The inset plot 
 displays the zoom in on the strong spectral feature which indicates periodicity of 33$\pm$4 periods.}
 \label{1741_quasi_periodicity}
\end{figure}

As evident from Figure \ref{1741_spdisplay}, the single pulses from \psra\ can also clearly be seen to 
switch quasi--periodically between the null and burst states which can be investigated further by 
taking a Fourier transform of the pulse energy modulation. Such bunching of burst and null pulses will 
give rise to periodic feature in the on--pulse energy fluctuation spectra, especially at lower fluctuation frequencies. 
\cite{wse07} have also reported a low frequency excess in their fluctuation spectra for \psra, however, no measurements 
was reported with possible quasi-periodicity of burst bunching. In more recent study, \cite{bmm16} conducted observations of \psra\ and 
reported drifting periodicities at 325 MHz and 610 MHz. 
Among the reported drifting features, low frequency periodic components were also 
reported with 0.033$\pm$0.006 cycles/P$_1$ and 0.033$\pm$0.009 cycles/P$_1$
at 325 MHz and 610 MHz, respectively. These low frequency drifting features correspond to roughly 
30--period quasi--periodicity which is likely to come from the bunching of null and burst pulses.
For \psrb, no indication of quasi-periodicity were reported in the previous studies. 

In order to measure this quasi-periodicity for the burst bunches, 
all the burst pulses were marked as ones and all remaining pulses (nulls) were marked as zeros for both pulsars. 
This one-zero sequence indicate emission state of the pulsar at each period, 
as shown in Figure \ref{1741_onps}. For \psra, it should be noted that 
a few burst bunches exhibit a small number of nulls, lasting for one or two periods, inside them. 
Similarly, a few nulls between the burst bunches also exhibit single individual burst pulses. 
As we are interested in the quasi-periodicity of the burst bunches, these pulses were ignored in the one-zero sequence. 
A Fourier transform was carried out on this one--zero sequence to quantify 
the quasi--periodic pattern for both pulsars. As suggested by \cite{hr07, hr09}, such technique helps in eliminating 
the pulse-to-pulse amplitude modulations and points toward exclusive modulation of the burst bunching. 
Figure \ref{1741_fft} shows the spectrum with a clear quasi-periodic feature around 33$\pm$4 periods for \psra. 
This periodicity also matches with the bunches seen in the null length and burst length histograms 
shown in Figure \ref{NL_BL_1741}. Thus, these measurements confirm that \psra\ exhibits 
clear quasi-periodic burst bunches. We did not find any strong feature in the similar spectrum for \psrb\ (not shown here). 
It is also likely that the absence of any periodic pattern in \psrb\ could be due to 
large NF in our comparatively small number of observed pulses.

\section{Pulse Energy modulations}
\label{Drifting_1741}
\label{drifting_sect}

\begin{figure}
 \centering
 \subfigure[]{ 
 \includegraphics[angle=0,bb=0 0 914 751,scale=0.23]{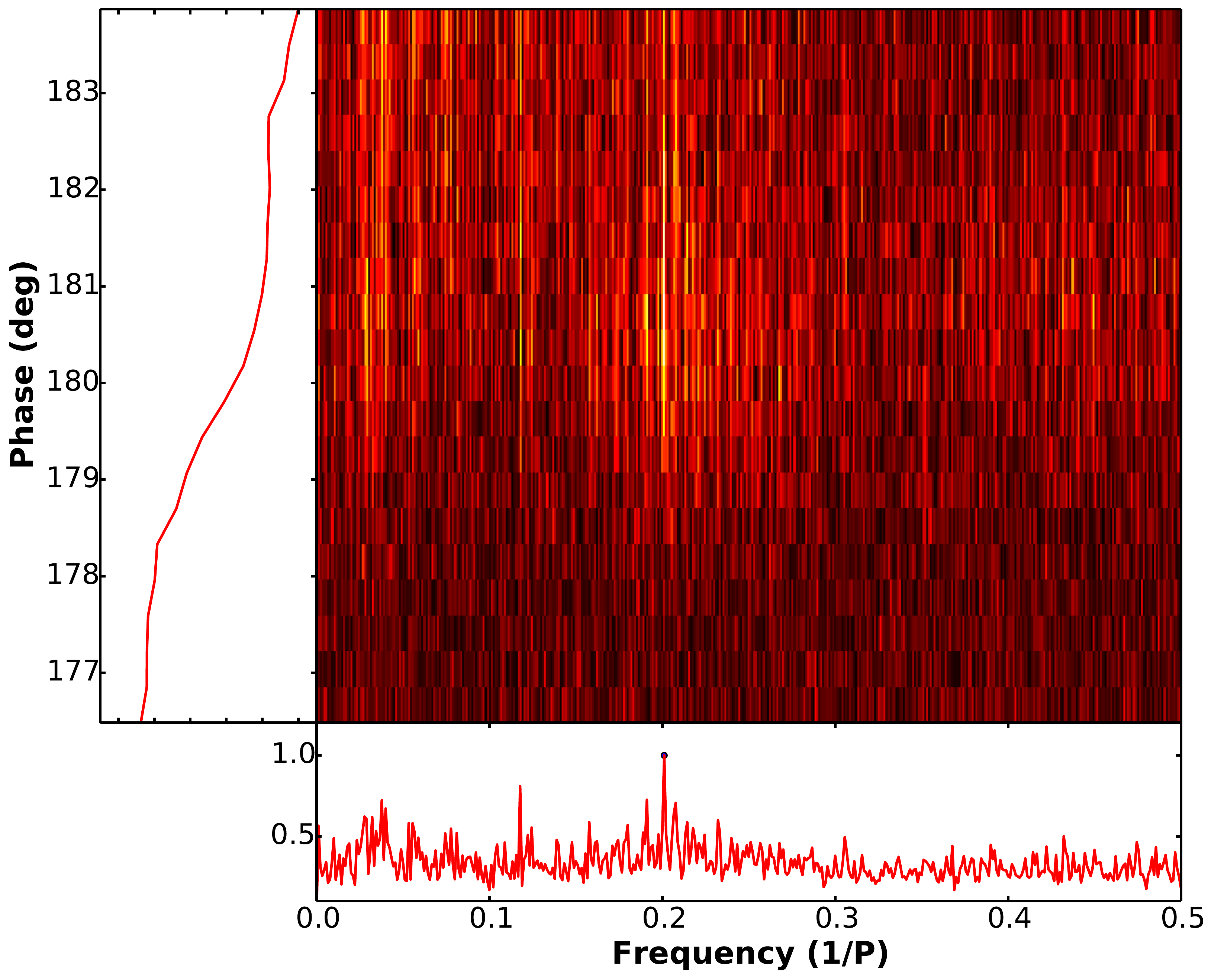}
 \label{1741_lrf_lead}
 }
 \subfigure[]{ 
 \includegraphics[angle=0,bb=0 0 914 751,scale=0.23]{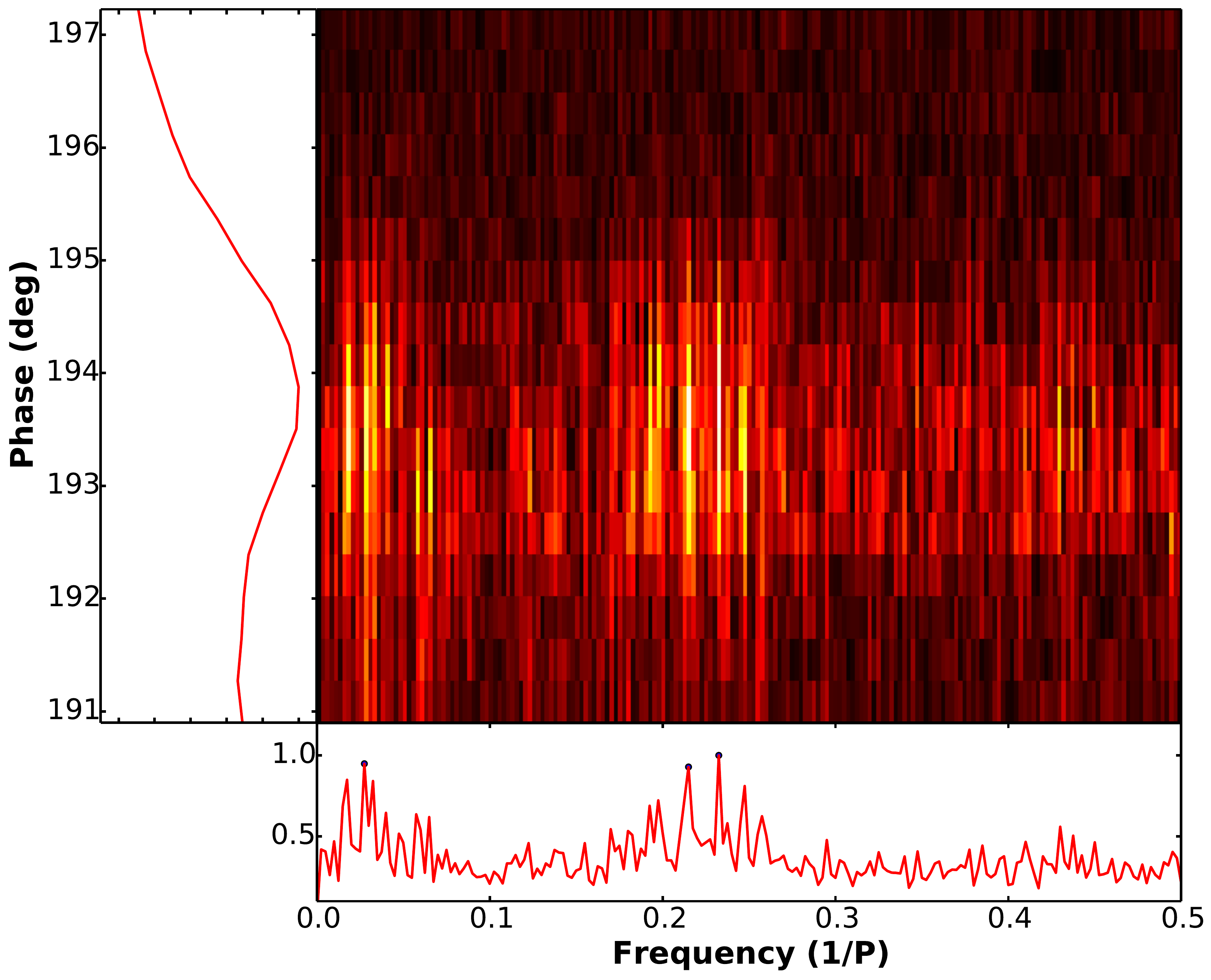}
 \label{1741_lrf_trail}
 }
 \caption{Longitude resolved fluctuations (LRFs) spectra of (a) leading 
 component, and (b) trailing component of \psra\ at 625 MHz. 
 The abscissa is the frequency of the drifting periodicity 
 and the ordinate is the profile phase for all three LRFs. The intensity of the 
 drifting periodicities as a function of pulse phase is plotted in color contours. 
 The leading component shows a modulation periodicity of 4.9$\pm$0.1 periods while the trailing component 
 shows two distinct periodicities of around 4.6$\pm$0.2 and 4.3$\pm$0.1.
 We fitted Gaussian functions at the center of these spectral lines and measured their 
 widths to estimate errors with 90\% confidence limit on the observed lines in the spectra.}
 \label{1741_lrf_fig}
\end{figure}

\begin{figure}
 \centering
 \includegraphics[scale=0.2,angle=0,bb=0 0 902 759]{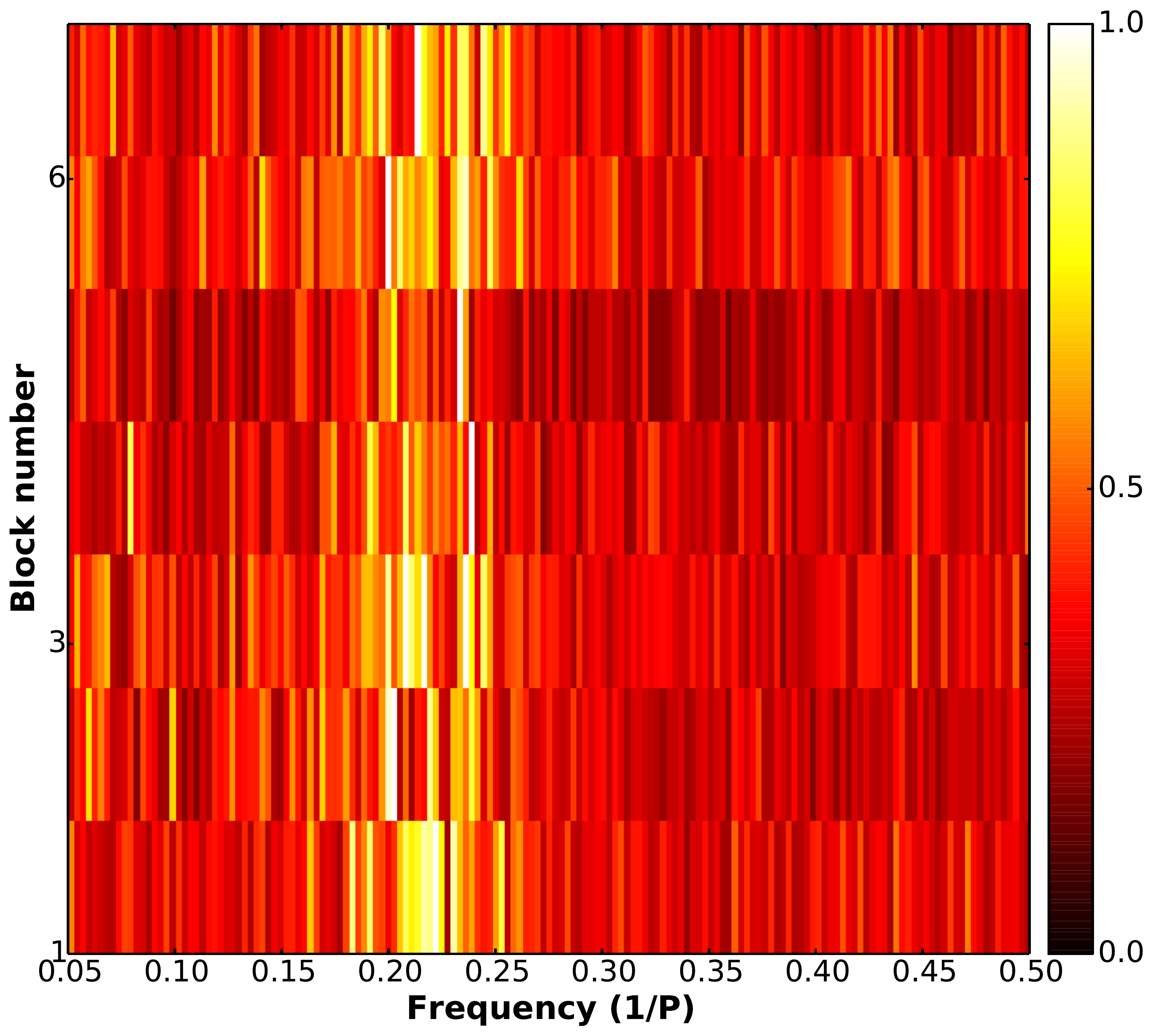}
 \caption{The LRF spectra of the trailing component of \psra\ at 325 MHz 
 from consecutive and overlapping block of 356 pulses with total 2206 single pulses. 
 Most of the blocks show diffuse periodicities without any significantly prominent 
 component.}	
 \label{1741bbb}
\end{figure}

A visual inspection of single pulses from \psra\ (Figure \ref{1741_spdisplay})  
clearly shows driftbands in the trailing component while the leading component 
does not show such distinctive driftbands. This is consistent with the reported 
behavior by \cite{wes06,wse07} at 1420 MHz and 325 MHz, respectively. We are 
unable to find any driftbands in the leading component, which indicates that 
any drifting periodicity seen in the component is likely to originate 
from possible amplitude modulation. 
Drifting in both components of \psrb\ is evident from 
Figure \ref{1840_spdisplay} but has never been reported in any previous studies. 
Although driftbands are prominent for both pulsars, the large NFs make the measurement 
of the drifting periodicity a challenging task. In order to estimate P$_3$, we measured 
the longitude-resolved fluctuation spectra (LRFS) for both pulsars. Collectively, the 
LRFS represents a Fourier transform computed over each individual phase bin, and can be 
represented in a 2-Dimensional contour plot along with average fluctuations taken over both axes. 
The pulsar period is first divided into 1024 phase bins for a given pulsar. 
A pulse--to--pulse fluctuation time--series is obtained for individual phase bin and  
a discrete Fourier transform is carried out over each of them to identify the fluctuation periodicities. 

Figure \ref{1741_lrf_fig} shows the spectra obtained for the leading and the trailing components of \psra. 
It should be noted that as \psra\ exhibit quasi-periodic fluctuation of around 33 periods due to burst bunches 
(see section \ref{qperiodicity_section}), thus the spectra obtained over a sequence containing nulls will be dominated by this quasi-periodic  
feature, as reported by earlier studies \cite[]{wse07,bmm16}. Such low frequency excess can be seen for both 
the components in Figure \ref{1741_lrf_fig}. The LRFS for the individual component suggests an interesting drifting behavior for this pulsar.
The leading component shows a significant peak around 4.9$\pm$0.1 periods along with its harmonics. 
The trailing component, which has width of around 6$^\circ\pm$0.5$^\circ$, exhibits different degrees of periodicities with a prominent 
and relatively narrow peak around 4.3$\pm$0.1 periods, and a relatively broader peak around 4.6$\pm$0.2 periods. 
These dual periodicities are intriguing and require a closer and detail look at the 
drifting behavior in the trailing component. It can be speculated that these periodicities 
are not likely to be similar for all burst pulses. As mentioned in Section \ref{intro}, 
a few pulsars exhibit time dependent drifting periodicities and switch between different drift modes. 
In order to scrutinize these drift modes, we computed a time-dependent LRFS for the trailing component. 
We first divided the observed sequence of 1206 pulses into 7 overlapping blocks, with each block consisting of 356 pulses. 
We then measured LRFS from each of the consecutive and overlapping blocks.
The result is shown in Figure \ref{1741bbb}, which clearly exhibits broad and diffuse spectra for 
all the blocks implying similar drifting features in each block without any significant differences. 
It should be noted that P$_3$ measured from the LRFS are averaged values which requires substantial 
number of periods with similar repeating pattern. 
As we are seeing multiple P$_3$ peaks for each block, 
it is likely that drifting might be changing during the course of a single block of 356 periods considered here.
Such rapid changes in the drifting periodicities, occurring on a smaller scale, might be possible 
to visually see from the driftband slopes. We will explore a detailed analysis of the drift 
band slope changes in Section \ref{drift_rate_sect}.

\begin{figure}
 \subfigure[]{
 \centering
 \includegraphics[angle=0,bb=0 0 938 759,scale=0.23,keepaspectratio=true]{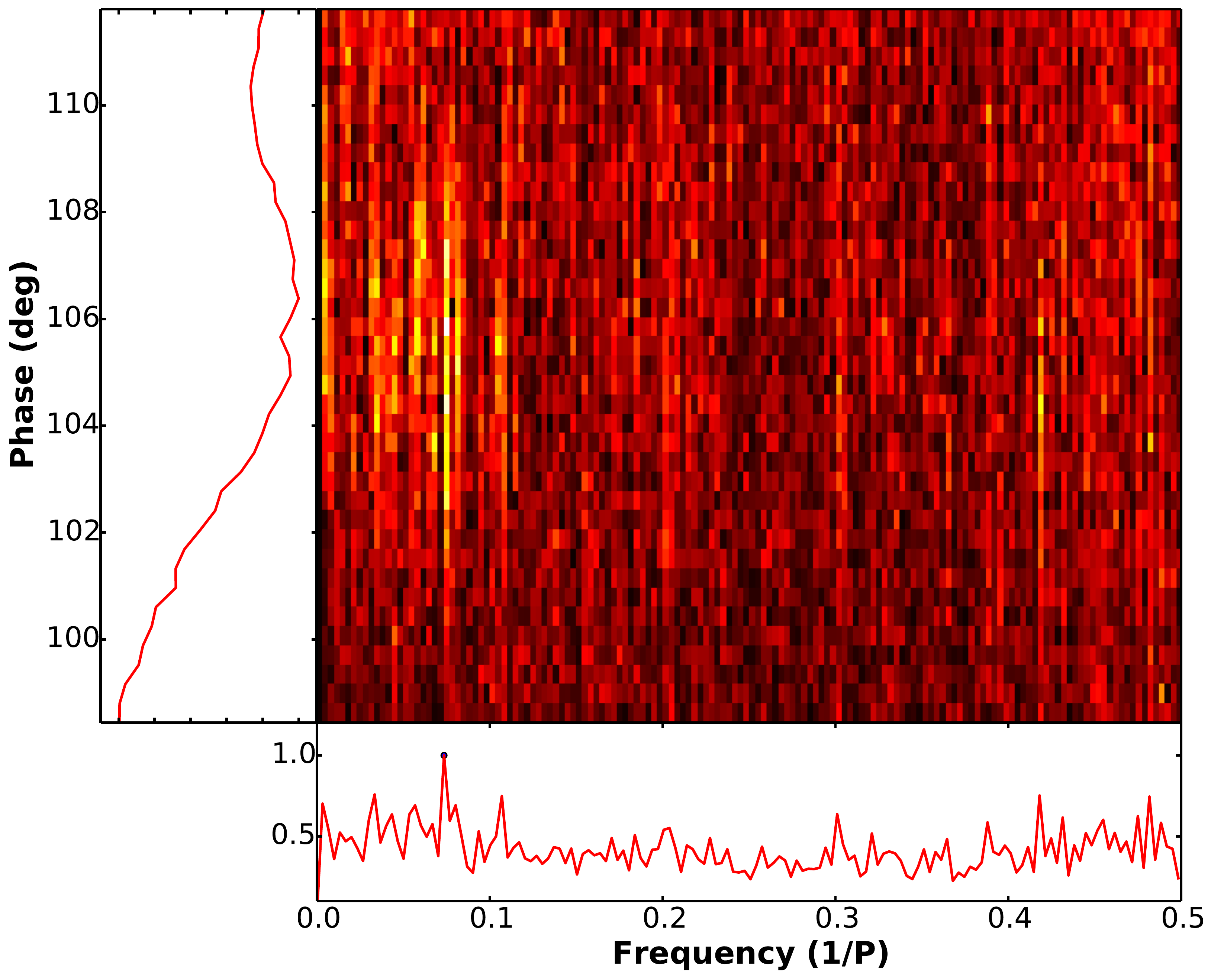}
 \label{1840_LRFlead}
 }
 \subfigure[]{
 \centering
 \includegraphics[bb=0 0 938 759,scale=0.23,angle=0]{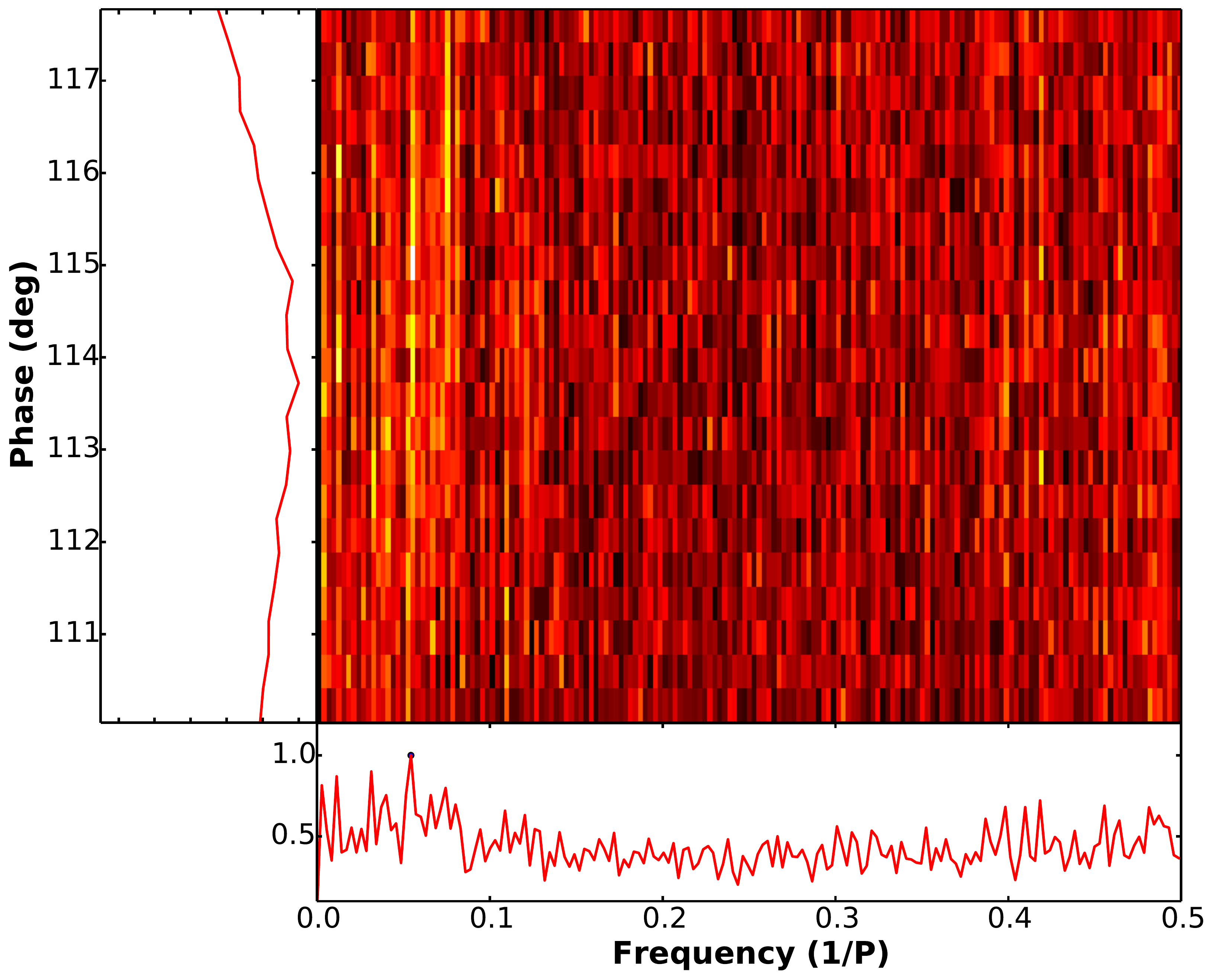}
 \label{1840_LRFtrail}
 }
 \caption{The longitude resolve fluctuation spectra of \psrb\ at 625 MHz for (a) the leading 
 component and (b) trailing component. The notation for each axis is similar to Figure \ref{1741_lrf_fig}. 
 The leading component exhibits drifting of around 13.5$\pm$0.7 periods while the trailing component 
 exhibits longer periodicity of around 18$\pm$1 periods.
 As this pulsar exhibits longer sequence of null pulses, we were only able to identify a 
 single sequence of 350 consecutive pulses which has relatively lower number of intervening long nulls. 
 However, it should also be noted that these measurements are likely to be
 influenced by these intervening nulls.}
 \label{1840_LRF}
 \end{figure}

For \psrb, the obtained LRFS for the leading and trailing components are shown in 
Figures \ref{1840_LRFlead} and \ref{1840_LRFtrail}, respectively. 
The leading component has width of around 11$\pm$0.5$^\circ$ and shows relatively stronger 
and broader peak of around 13.5$\pm$0.7 periods, as seen in Figure \ref{1840_LRFlead}. 
A visual examination of the observed single pulses reveals that this periodicity is not 
stable and exhibits a range of driftband shapes. These changes are further 
quantified in Section \ref{drift_rate_sect}. The trailing component of this pulsar has width of around 
9$^\circ\pm$0.5$^\circ$ and an estimated P$_3$ of around 18$\pm$1 periods.
Moreover, it can be seen from Figure \ref{1840_LRFtrail} 
that the significance of P$_3$ is relatively low compared to the P$_3$ of the leading component. 
It should be noted that leading component shows slightly lower P$_3$ compared to the trailing component. 
Such slight differences in P$_3$ periodicity between different components are seen in a few pulsars in 
earlier studies. For example, PSR B0826--34 showed different P$_3$ for different profile components \cite[]{bmh+85}. 
However, it should be noted that as the pulsar exhibits longer nulling instances, our P$_3$ measurements, which were made from 
only a single sequence of around 350 consecutive pulses, is likely to be hampered by these intervening nulls.  
A more detailed comparison using the driftband slopes is required to scrutinize
these differences of drifting between both components of \psrb.

\section{Driftband slopes}
\label{drift_rate_sect}
\label{drift_rate_sect}
\begin{figure}
 \centering
 \includegraphics[scale=0.3,angle=0,bb=0 0 630 970]{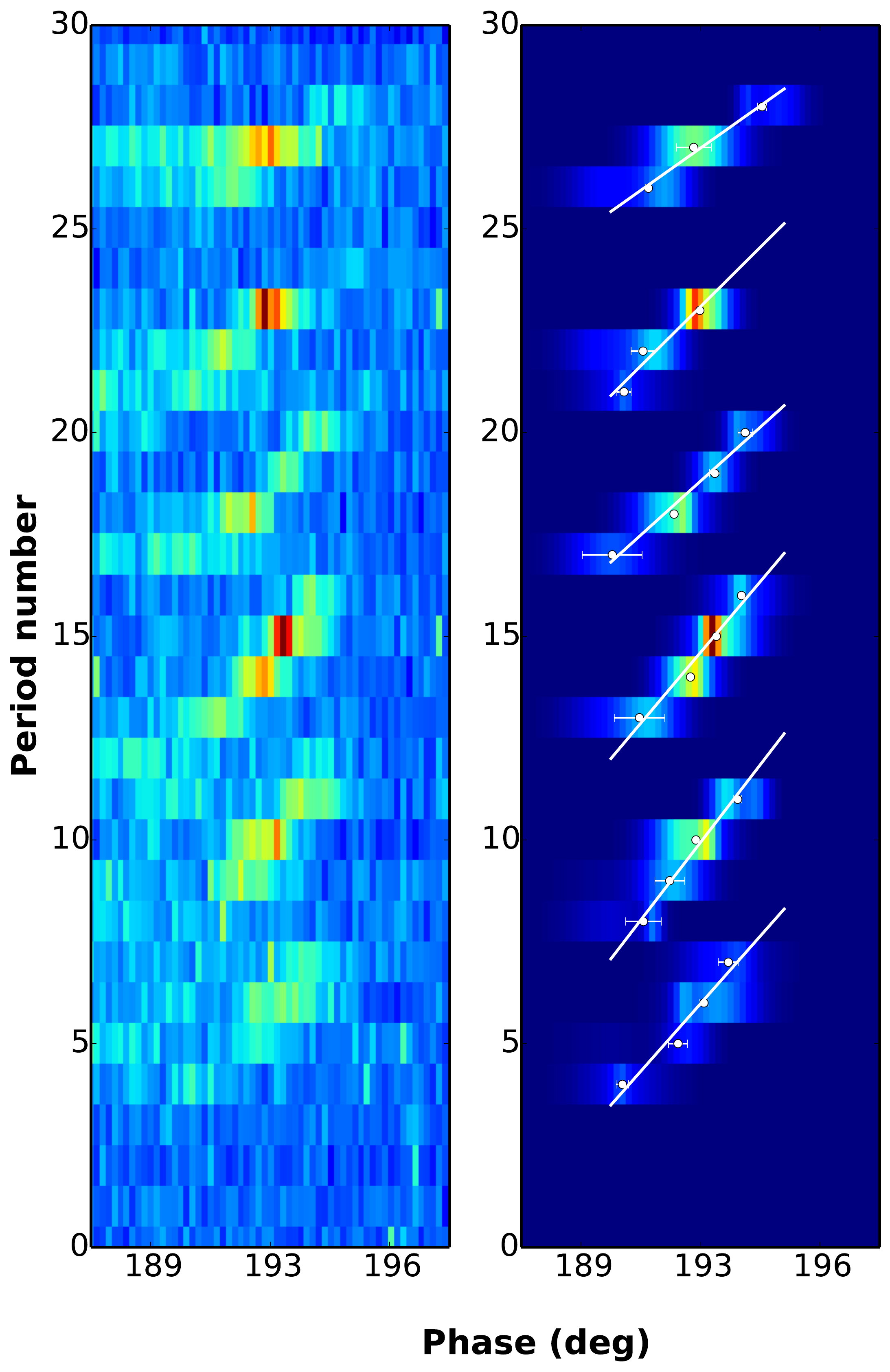}
 \caption{A section of {\br the} observed drifting in the trailing component of \psra\ at 625 MHz 
 is shown in the left plot. The corresponding Gaussian modeled components, along with 
 the best linear fits (white straight lines) for each driftband, {\br are} shown in the right plot.}
 \label{drifting_fit_1741}
\end{figure}
\begin{figure}
 \subfigure[]{
   \centering
   \includegraphics[scale=0.2,bb=0 0 925 773]{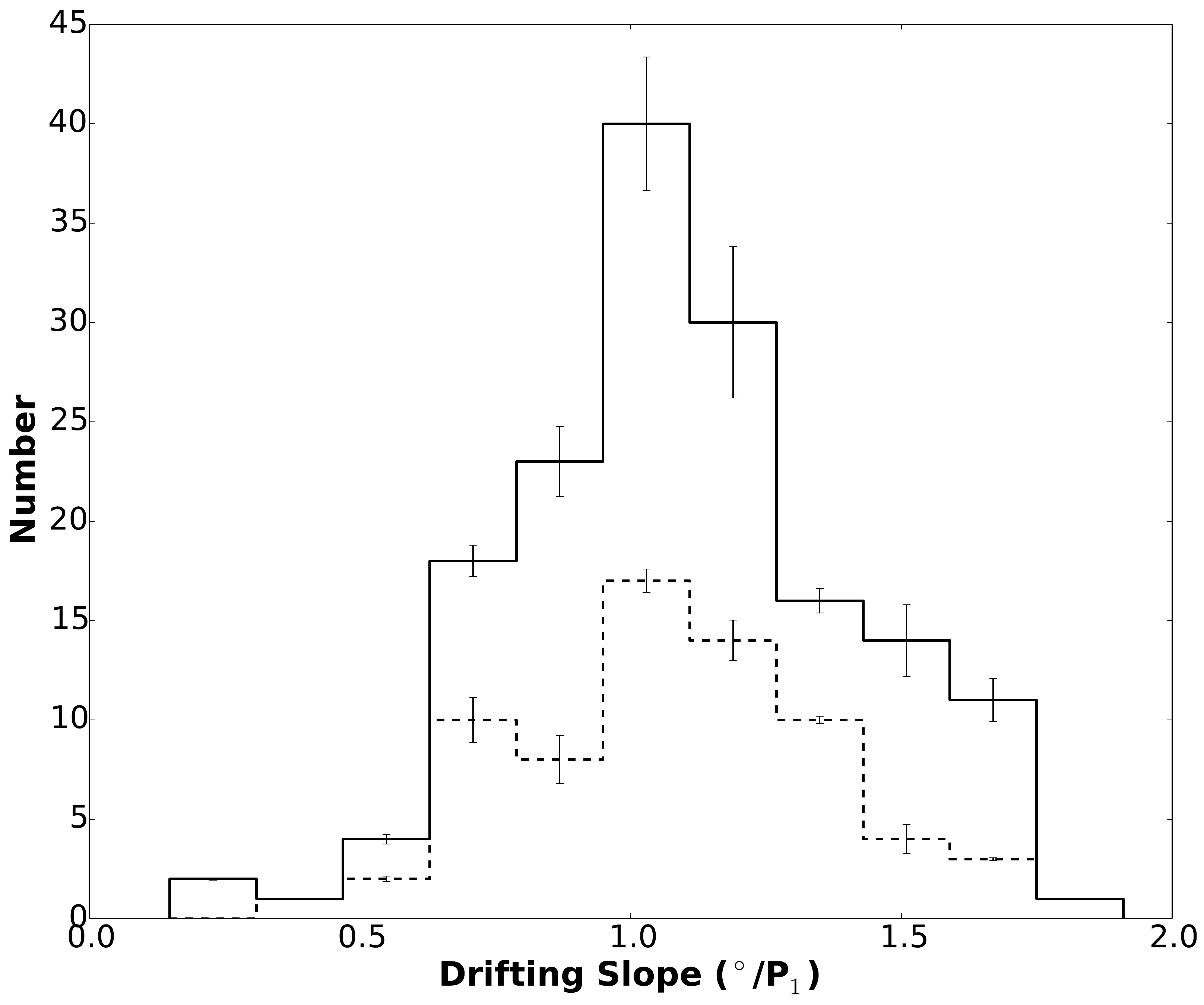}
   \label{drift_rate_1741}
   }
 \subfigure[]{  
   \centering
   \includegraphics[scale=0.2,bb=0 0 925 773]{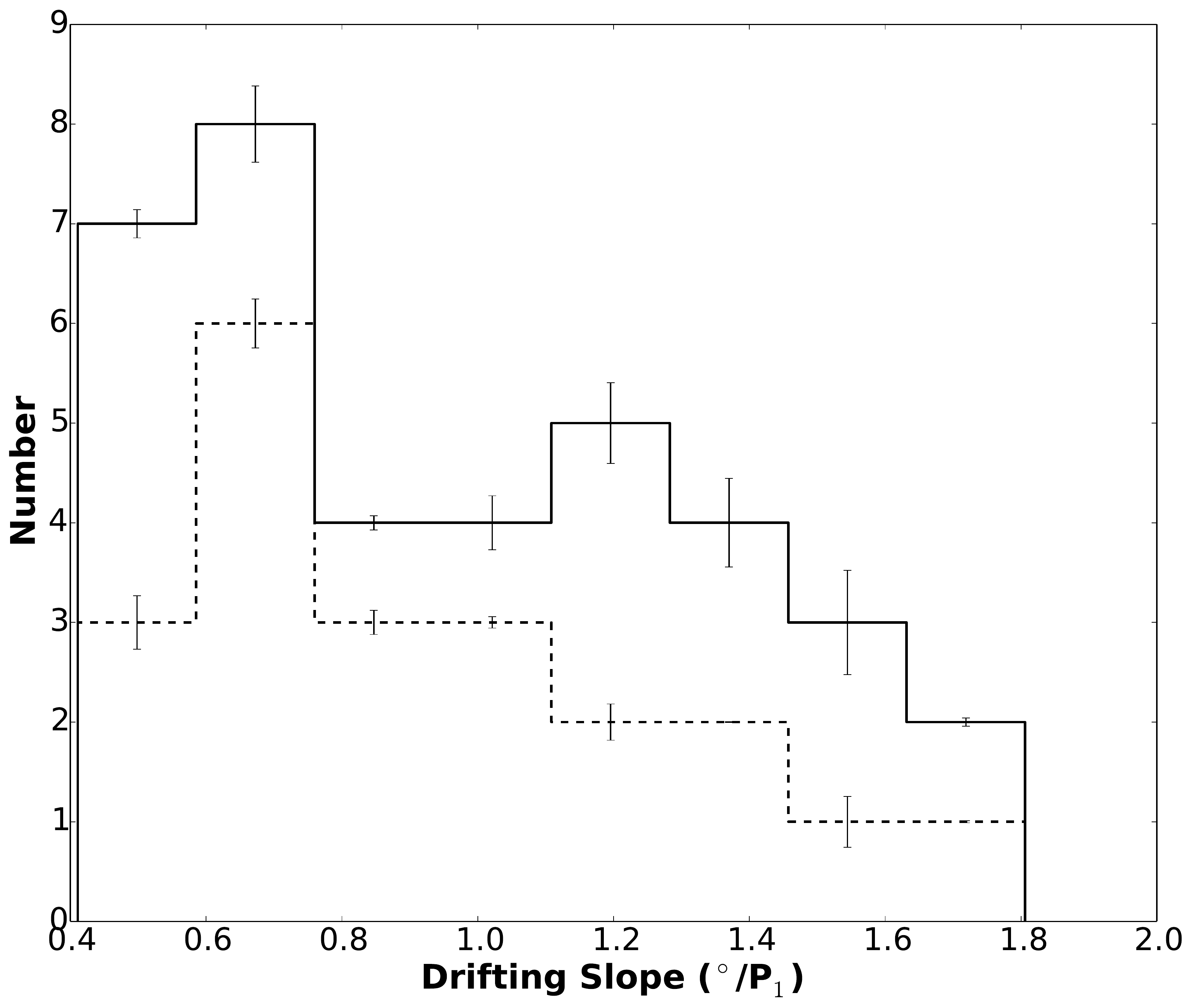}
   \label{drift_rate_1840}
  }  
 \subfigure[]{
 \centering
 \includegraphics[scale=0.2,bb=0 0 925 773]{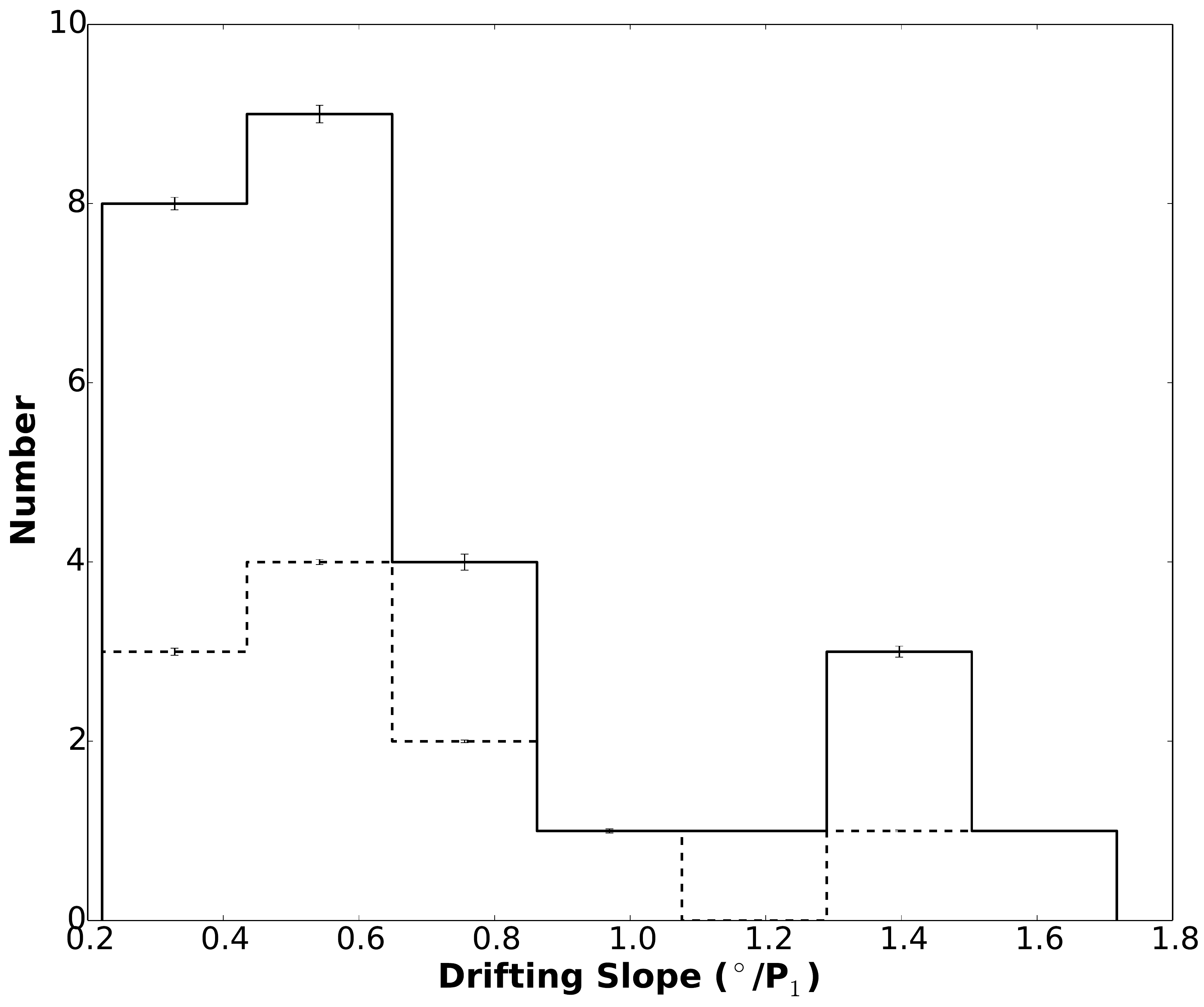}
 \label{drift_rate_1840_trailing}
 }
  \caption{Histograms of measured drift-rates in (a) trailing component of the \psra, 
  (b) leading, and (c) trailing components of \psrb\ at 625 MHz. The histograms with dotted line represent the 
  drift-rates measured from the drift-bands around nulls. The error bars are obtained for 
  each bin incorporating errors from the measured D in corresponding histogram bin. 
  Both pulsars show no significant difference between the average drift-rates and the drift-rates measure around nulls.}
  \label{drift_rate_hist_fig}
\end{figure}

The spread in the driftband slope (hereafter drift--rate; D) can point towards possible 
drift mode changes which can occur on a smaller time--scale compared to number periods required to measure 
the average P$_3$. If there are stable and distinct drift modes in a given pulsar, the measurements 
of D are likely to show some clustering corresponding to these modes.
It should be noted {\br that} as D is a ratio between two quantities (D~$=~$P$_2$/P$_3$), there 
is an inherent degeneracy regarding which parameter dominates that causes D to change. 
In other words, a similar fluctuations in sub--pulse separation (P$_2$) can also make corresponding 
changes in the D. However, \cite{wse07} have carried out study of a few hundred pulsars and concluded 
that P$_2$ does not change during different drift modes. Moreover, a few earlier studies have also 
suggested stability of P$_2$ \cite[]{htt70,tmh75,la83} in different pulsars.  
Thus, treating P$_2$ as a constant for both pulsars allows us to assume that 
any significantly different values exhibited in D corresponds to different P$_3$ measurements. 
We have used this same assumption in Sections \ref{drift_phase_memory_sect} and \ref{discussion_and_conclusion_sect}. 

In order to measure D with high significance, we modeled each single pulse from a given component $-$ trailing for \psra\ and both components  
for \psrb\ $-$ using a Gaussian function for both pulsars. 
Such modeling has been carried out for many pulsars in earlier studies \cite[]{la83,vj97,vj99,vkr+02}. 
It provides unique smoothness to the true shape of the single pulse components, which is helpful in 
determining the drift--rate and its associated changes. Each component of every burst pulse was carefully modeled 
with a Gaussian function in order to keep the reduce Chi-square close to unity. 
Then, each fitted model sub--pulse was carefully examined visually to verify its representation of 
the true shape of the component under consideration. This modeling gives a 
center point for each sub--pulse which can be traced by fitting a straight line across a given driftband  
to determine the corresponding D. Figure \ref{drifting_fit_1741} shows an example of several driftbands 
in the trailing component for a section of burst bunch from \psra. To determine the slope of these lines, 
individual errors in the estimation of modeled centers were also incorporated. Thus it was possible to 
obtain a more accurate estimation (with acceptable reduced Chi--square) of the slope with more stringent 
error only for driftbands with sufficient number of sub--pulses. For weak sub--pulses, the error bars on 
the estimated centers were relatively larger which propagated during the estimation of the driftband slopes,  
resulting in larger uncertainties. Figure \ref{drift_rate_hist_fig} shows distribution of D obtained 
for both pulsars for corresponding components. 

For \psra, distribution of the measured drift--rates (shown in Figure \ref{drift_rate_1741}) 
displays a large spread corresponding to driftbands with different slopes.
The mean value of the measured D ($\hat{D}$) is around 1.1$^{\circ}\pm$0.2$^\circ/P_1$, which 
was obtained from around 160 observed driftbands in the trailing component of \psra. 
It should be noted that the error bar on the measured D is large enough to produce different drifting 
periodicities seen in the LRFS given a relatively constant P$_2$. 
Although the D shows large spreads, we do not notice any significant clustering 
that could point to the presence of different drift modes. Thus, we did not identify any drift-modes in this pulsar 
which could explain the different P$_3$ periodicities seen in the LRFS. It can be speculated that, the drifting is very sporadic and  
it changes irregularly without clearly producing distinguishable drift modes. Such changes can also be seen towards the 
last pulses in Figure \ref{drifting_fit_1741} for \psra. 

Figures \ref{drift_rate_1840} and \ref{drift_rate_1840_trailing} show distributions of D for \psrb\ which also exhibit significant spreads
for both components. Since \psrb\ exhibits large fraction of null pulses, we identify only 
37 and 27 significant driftbands for measurement of their slopes with good Chi-square fits in the leading and 
trailing components, respectively. 
From Figure \ref{drift_rate_1840} it can be speculated that the D has bi--modal distributions as indicated by 
the two peaks. The drift rate of the trailing component, shown in Figure \ref{drift_rate_1840_trailing}, 
does show similar drift rates to the leading component although the number of {\itshape good} 
driftbands are relatively lesser. In order to scrutinize this bi--modality in D for both components, 
we carried out {\itshape Hartigans' Dip Test for Unimodality} proposed by \cite{hh85}. The test 
assumes a null hypothesis of unimodal distribution and 
provides a {\itshape p-value} for a given input samples. We found that the {\itshape p-value} for unimodal distribution 
hypothesis is 93\% and 95\% for the leading and trailing components, respectively, 
which clearly suggest that the distribution of D is most likely to be unimodal for both components.
It should be noted that our sample size was not sufficient enough (specially for the trailing component) 
to clearly claim bi--modality in Figures \ref{drift_rate_1840} and \ref{drift_rate_1840_trailing} with high significance. We simulated 
around 20,000 D values for each components using Monte-Carlo technique and carried out the {\itshape Dip test} 
again but did not find any significant changes in the {\itshape p-value}. 
Thus, considering unimodal distribution of D we estimated average $\hat{D_{l}}$ 
of around 0.9$^{\circ}\pm$0.4$^\circ/P_1$ and $\hat{D_{t}}$ of around  0.7$^{\circ}\pm$0.4$^\circ/P_1$, for the leading and trailing components, respectively. 
The $\hat{D_{t}}$ is comparable to $\hat{D_{l}}$ within error bars, which 
suggests that both components exhibit similar drifting periodicity that was not 
possible to scrutinize from the measurements of P$_3$ in the individual component due to large nulling instances 
(see Section \ref{drifting_sect}). Thus, it should be noted that such large spread in drift--rates can 
contribute to the broad and different P$_3$ periodicities seen in both components of \psrb\ given a constant P$_2$.

Furthermore, we did not find any correlation between D and pulse-to-pulse intensity fluctuations to hint at
any association between them.
As both pulsars clearly show large fraction of nulls, it is likely that these nulls might be 
interacting significantly with these drift rates. This is discussed further in the next section. 

\subsection{Null associated drift rate changes}
\label{null_induce_D_sect}
Nulls are known to cause significant changes in the drifting behavior of the adjacent burst pulses 
for a few pulsars \cite[]{la83,dchr86,vj97,jv00,vkr+02}. 
The burst pulses occurring before or after a null phase are known to show noticeable changes
in the drifting slopes like that in PSR B0809+74  \cite[]{la83,vkr+02} or it could 
also act as a trigger to cause a pulsar to switch between different drift-modes 
like in the case of PSR J1727$-$2739 \cite[]{wwy+16}.
In order to investigate any changes in the drift slopes caused by nulls, 
we separated all the driftbands occurring before and after the null states 
for both pulsars. 

For \psra, driftbands before the onset of nulls showed mean D ($\hat{D}_{b}$) of around 1.12$^{\circ}\pm$ 0.2$^\circ/P_1$ 
while the driftbands after the nulls showed mean D ($\hat{D}_{a}$) of around 1.05$^{\circ}\pm$0.2$^\circ/P_1$.  
These are not significantly different from the $\hat{D}$ measured from all driftbands. 
Similarly, for \psrb\ we separated all the driftbands around the nulls for both components.
We only used driftbands which had sufficient S/N and provided good fit to 
the straight line tracing of its drifting direction. We found $\hat{D}_{la}$ for the
driftbands after the null states to be around 0.97$^{\circ}\pm$0.21$^\circ/P_1$ and $\hat{D}_{lb}$ 
for the driftbands before the onset of nulls is 0.93$^{\circ}\pm$0.27$^\circ/P_1$ for the leading component.
These measurements match with the average value ($\hat{D}_l$) within the error bars. 
For the trailing component, we found $\hat{D}_{tb}$ and $\hat{D}_{ta}$ to be around 0.73$^{\circ}\pm$0.54$^\circ/P_1$ 
and 0.64$^{\circ}\pm$0.21$^\circ/P_1$ for drift--rates before and after the nulls, respectively.
These are also not significantly different from the $\hat{D}_{t}$.
Figure \ref{drift_rate_hist_fig} shows a separate 
distribution of measured D around nulls for both pulsars. 
As the driftbands do not show significantly different distributions, 
we conclude that none of the components exhibit noticeable driftband slope changes influenced by nulls.

\begin{figure*}
 \centering
 \subfigure[]{
 \includegraphics[scale=0.65,bb=0 0 360 252]{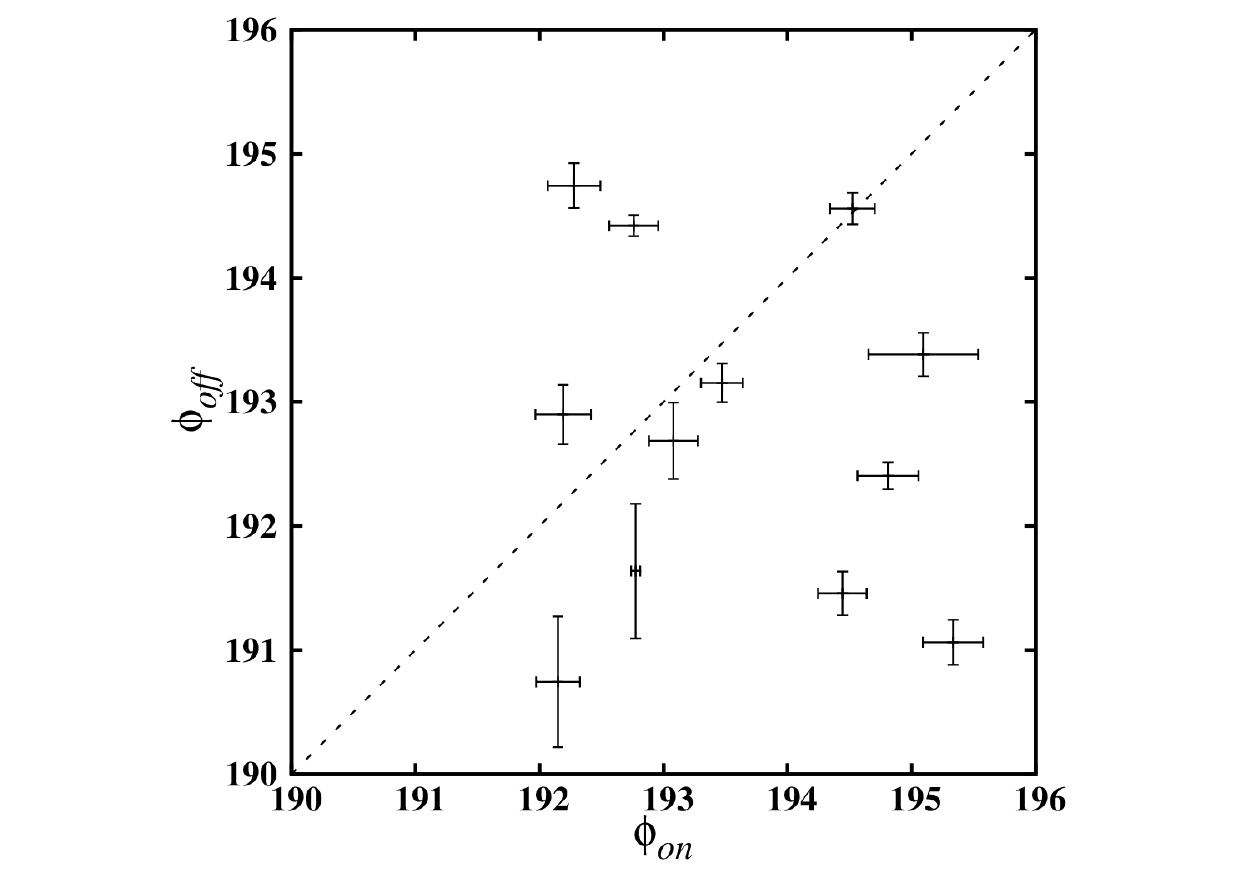}
 \label{turnoff_turnon}
 }
 \subfigure[]{
 \centering
 \includegraphics[bb=0 0 360 252,scale=0.65]{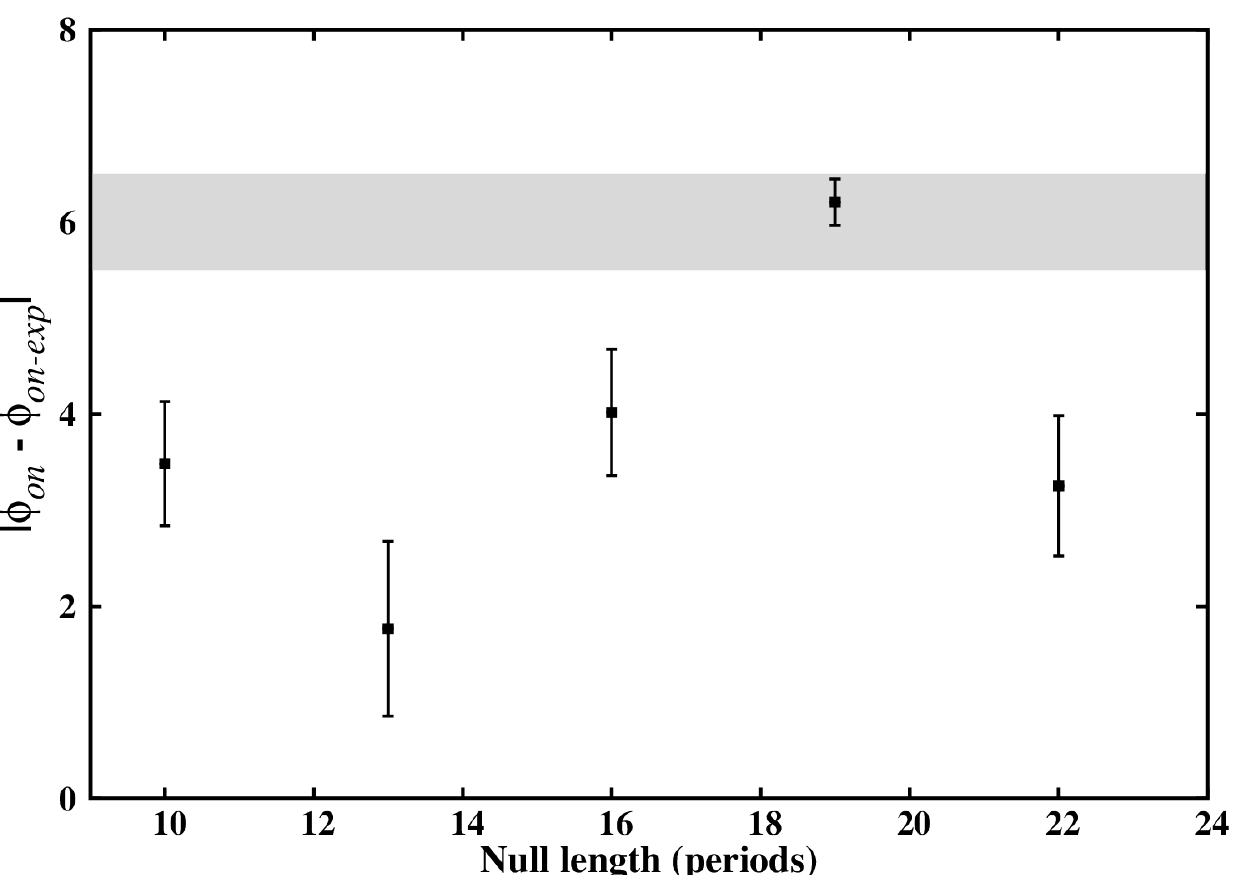}
 \label{stopage_drifting}
 }
 \caption{Comparison between the pulse phase across nulls for the trailing component 
 of \psra\ at 625 MHz. (a) The distribution of the turn--on phase versus the turn--off phase across nulls, and 
 (b) absolute phase differences across the nulls as a function of null lengths. 
 The shaded region in the right hand panel corresponds to the average width of the component with the errors.
 We binned the measured $\bigtriangleup{\phi}$ into bins of five null lengths, 
 shown here as filled squares corresponding to given null length bin, along with the corresponding error bars.}
\end{figure*} 

\section{Drifting phase memory across nulls}
\label{drift_phase_memory_sect}
In a very few nulling pulsars, drifting seems to retain some memory of its phase from before 
the pulsar goes into a null state. Such memory may provide important link in understanding the
true nature of the nulling phenomenon. We discuss two ways in which this phase memory across nulls 
can be scrutinized.

\subsection{Turn--on and Turn--off phase preference}
\label{turn_onoff_pref_sect}
\begin{figure*}
 \centering
 \subfigure[]{
 \includegraphics[width=6 cm,height=6.3 cm,angle=0,bb=0 0 616 655]{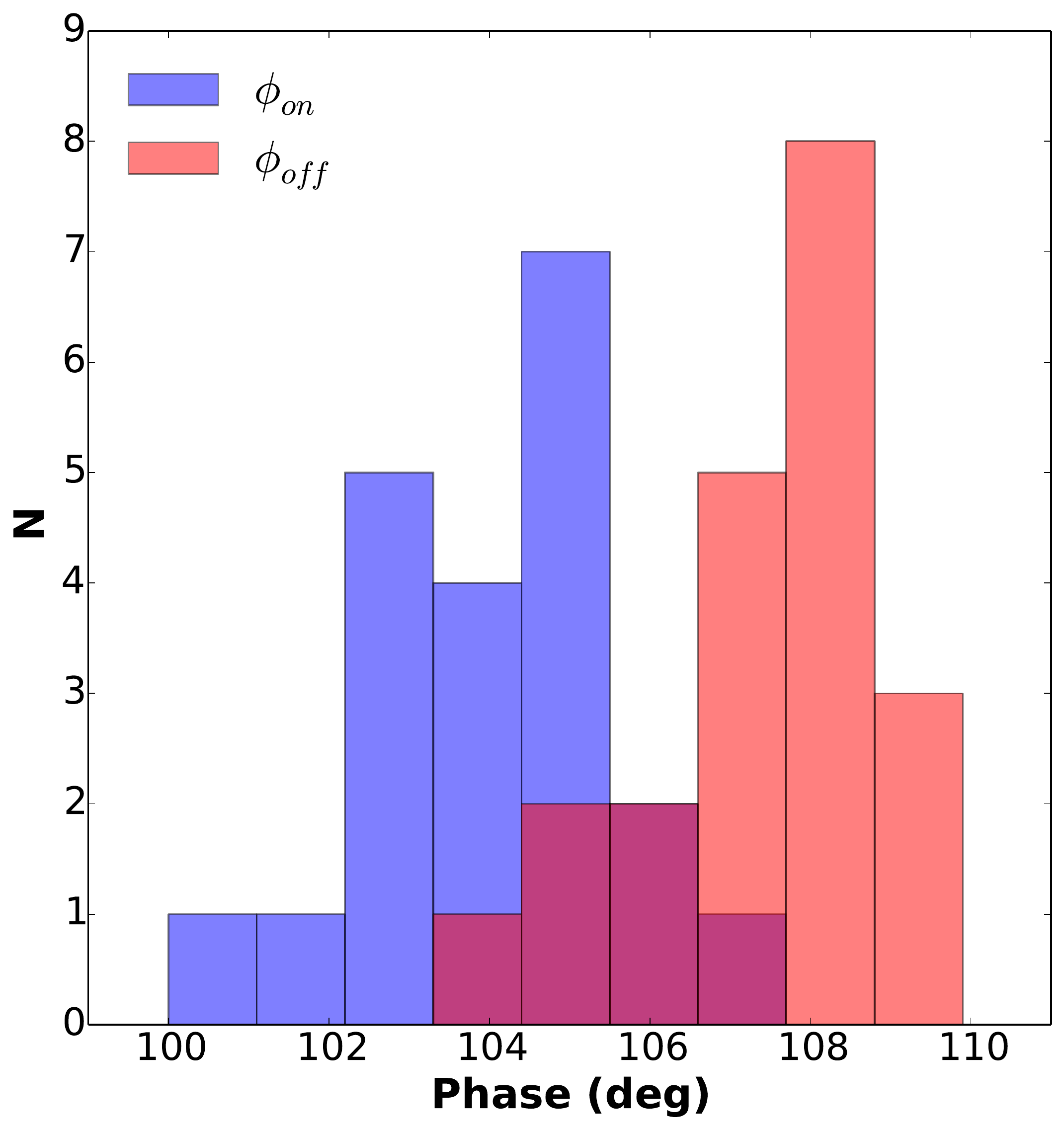}
 \label{Turn_off_Turn_on_phase_1840}
 }
 \centering
 \subfigure[]{
 
 \includegraphics[scale=0.7,angle=0,bb=0 0 360 252]{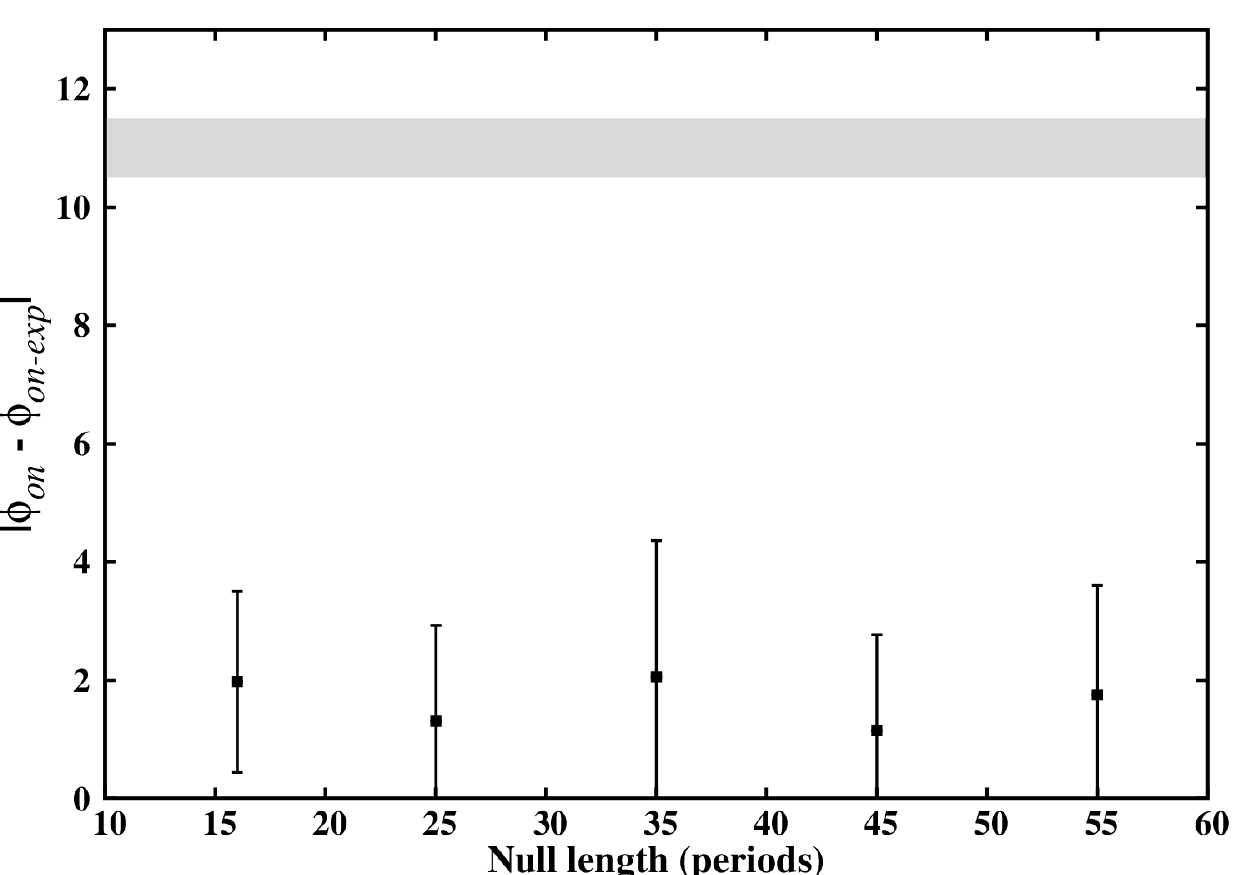}
 \label{Stopage_drifting_1840}
 }
 
 \caption{The relation between the turn--on and turn--off phases for the leading component of \psrb\ at 625 MHz. 
 (a) Histograms of measured distribution of $\phi_{on}$ and $\phi_{off}$ which clearly show very distinct distribution suggesting peculiar 
 preferences of burst pulse phase around nulls. (b) Absolute phase differences ($\bigtriangleup{\phi}$) 
 as a function of null lengths. The measured null--lengths 
 were binned into seven bins and average $\bigtriangleup{\phi}$ was measured for each bin.
 The shaded region in the right hand panel corresponds to the average width of the leading component with the errors.}
 \label{Turn_off_Turn_on_phase_1840_full}
\end{figure*}

\begin{figure*}
 \centering
 \subfigure[]{
 \includegraphics[width=6 cm,height=6.3 cm,angle=0,bb=0 0 616 655]{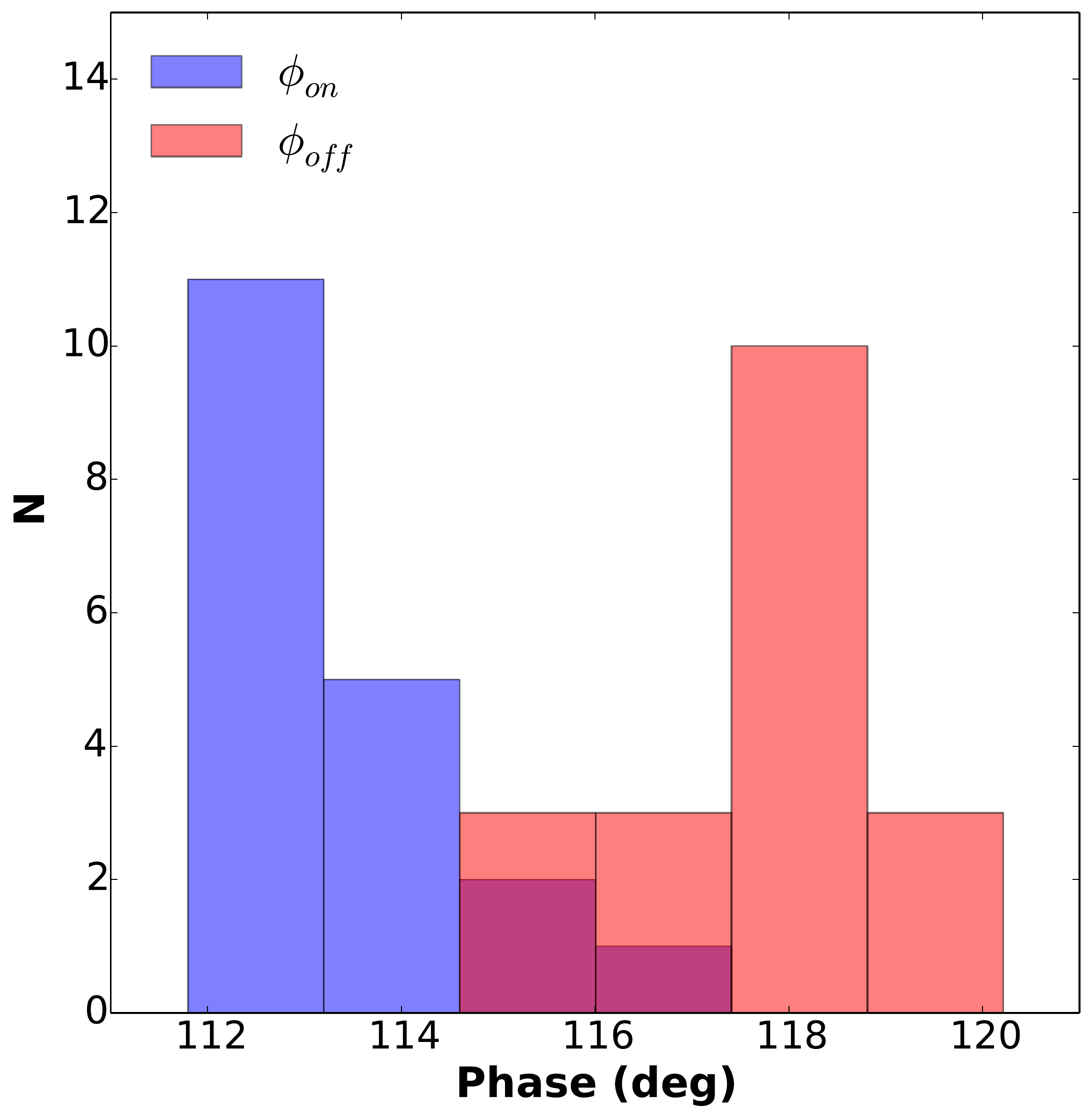}
 \label{Turn_off_Turn_on_phase_1840_trail_fig}
 }
 \subfigure[]{
  \includegraphics[scale=0.7,angle=0,bb=0 0 360 252]{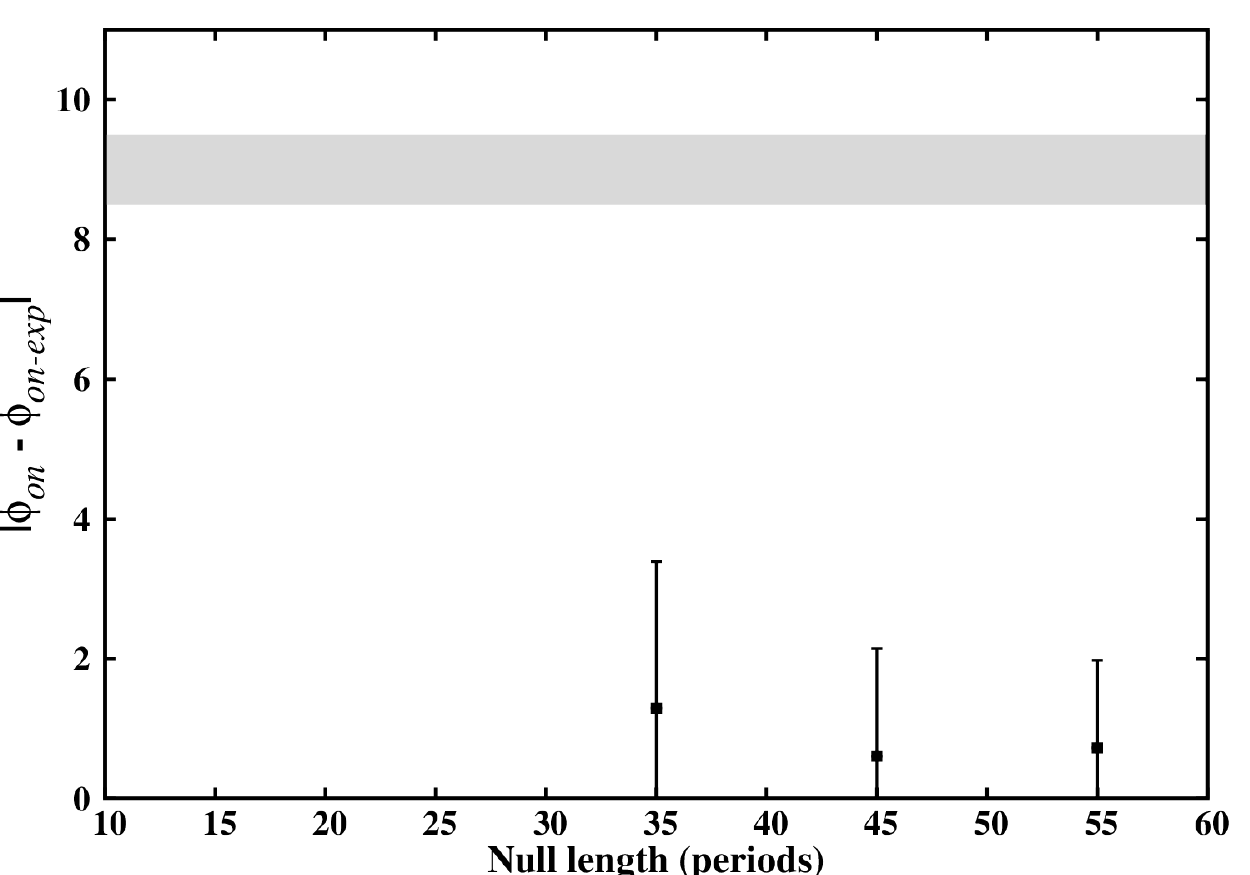}
 \label{Stopage_drifting_1840_trail}
 }
 \caption{Similar relation between the turn--on and turn--off phases for the trailing component of \psrb\ at 625 MHz.
 (a) Histograms of measured distribution of $\phi_{on}$ and $\phi_{off}$ for the trailing 
 component of \psrb\ at 625 MHz. The distribution clearly shows very distinct distribution suggesting peculiar 
 preferences of burst pulse phase around nulls also occur in the trailing component.
 (b) Absolute phase differences ($\bigtriangleup{\phi}$) as a function of null lengths. The measured null--lengths 
 were binned into three bins and average $\bigtriangleup{\phi}$ was measured for each bin. 
 The shaded region in the right hand panel corresponds to the average width of the trailing component with the errors.}
 \end{figure*}

\cite{jv00} suggested a comparison between the turn--off phase and the turn--on phase 
to investigate phase memory across nulls. In our case, these phases correspond to 
the center of each fitted Gaussian components of the burst pulses 
at the null--to--burst and burst--to--null transition instances. 
If a pulsar prefers particular phase during the transitions, last burst pulse before 
the null and first burst pulse after the null should exhibit distribution of phases significantly 
different from the average center for a given component. We obtained distributions of the 
turn--off phases ($\phi_{off}$) and turn--on phases ($\phi_{on}$) for the corresponding components in both pulsars. 
It should be noted that when calculating a first pulse after the null and last pulse before the null, we
only consider emission in the component under consideration. Both components in both pulsars tend to show simultaneous  
transition instances with a very few exceptions and we assume that our outcome is not likely to get much 
influenced by these small discrepancies.

We did not find any preferences for the turn--off and turn--on phases 
for the trailing component of \psra\ (see Figure \ref{turnoff_turnon}). 
Thus, it can be speculated that \psra\ does not prefer any particular turn--on or 
turn--off phase around the null instances for its trailing component. 
We should point out that this non--correlation between the turn--on and turn--off phases can also 
be due to shorter drifting periodicity. 
Given a P$_3$ of around 4.3 and 4.6 periods, 
in a complete cycle of a driftband we are only 
going to sample the sub-pulse phase 4 to 5 times. 
This can influence a correct measurement 
of turn--on and turn--off phases for around 25\% of the time either way. 
Since we only study this behavior for only 12 transitions, our sample size is 
limited to rule it out entirely.

\psrb\ was seen to have a very strong preference for turn--on and 
turn--off phases for both components. Figures \ref{Turn_off_Turn_on_phase_1840} 
and \ref{Turn_off_Turn_on_phase_1840_trail_fig} highlight preferred localization of $\phi_{on}$ and $\phi_{off}$ from the histogram 
of their measured distributions for leading and trailing components, respectively. 
The average turn--on phase for the leading component was found to be 
around 103.9$\pm$0.2 which is significantly different from the average turn--off phase of around 106.9$\pm$0.1. 
Similarly, the trailing component showed average turn--on phase of around 113.2$\pm$0.3 which is also noticeably 
different from the turn--off phase of around 117.7$\pm$0.2.
A Kolmogrove-Smirnov (KS) test of comparison between these distributions for both components rejected the null hypothesis of similarity 
with 99.9\% probability (with KS-test {\itshape D} value of 0.76 for leading component and 0.84 for the trailing component), 
suggesting that these distributions are very distinct.

Thus, it is likely that both components of \psrb\ exhibit preferred localization for the 
turn--on and turn--off phases. In other words, it appears that nulls always start after completion of a full driftband 
and it is very likely to end with a beginning of a new driftband in either components. 
These results also suggest that typical burst durations are integer multiple of P$_3$ periodicities 
which can be seen from Figure \ref{NL_BL_1840} with peaks around 15, 30 and 60 periods. 
Also typical bursts with around 15 periods for the leading 
and trailing components can also be seen in Figures \ref{simulation_fit_fig2} and \ref{1840_spdisplay}. 
A few other examples of null instances along with preferred turn--on phase 
can also be seen from single pulses in Figure \ref{1840_spdisplay}. 
This has never been seen with such high significance in any pulsars to the best of our knowledge. 

\subsection{Drifting across nulls}
\label{drifting_during_null_sect}
In this section we attempt to scrutinize motion or displacement of sub-pulses during the null states. 
As an extreme scenario, if drifting completely halts during nulls and the pulsar retains its phase memory, 
the turn--on phase is likely to be similar to the turn--off phase. In order to test this scenario for \psra, 
which exhibits range of turn--on and turn--off phases, we measured a correlation co-efficient between them. 
If turn--on and turn--off phases have any association across nulls, we are likely to see significant amount 
of correlation between them. We only compared those transitions where burst pulses $-$ near transitions $-$ 
have sufficient S/N and phases can be determined with relatively smaller errors. 
Figure \ref{turnoff_turnon} shows relation between the $\phi_{off}$ and $\phi_{on}$ for 12 such null 
transitions in the trailing component of \psra. For a complete stoppage of drifting, 
$\phi_{off}$ and $\phi_{on}$ should exhibit linear trend. It is clear from this comparison 
that there is no significant correlation between $\phi_{off}$ and $\phi_{on}$ $-$ with 
cross-correlation coefficient less than 0.02. For \psrb, on the contrary, we observed 
completely anti-correlated $\phi_{off}$ and $\phi_{on}$ (seen from Figures \ref{Turn_off_Turn_on_phase_1840} 
and \ref{Turn_off_Turn_on_phase_1840_trail_fig}). 
Implication of which are discussed further below. 

Although nulling represents complete absence of any detectable emission, the charging and 
discharging of the polar gap may still be operating. If sub--pulses on the emission beam 
continue to drift $-$ with similar or different rotation speed $-$ during the null state, 
the difference between the turn--on and the turn--off phases can be predicted based on the 
length of nulls. Furthermore, it is also likely that a pulsar retains memory of {\br the} turn--off phase 
only for certain length of nulls as in the case of PSR B0031$-$07 where only short nulls 
were shown to have such memory \cite[]{jv00}. Thus, there are two scenarios that need be tested, 
(a) drifting continues at its regular speed or (b) drifting slows down or speeds up during nulls. 
If drifting continues at its regular speed, it would be possible to predict location of the expected 
turn--on phase ($\phi_{on-exp}$) from Equation \ref{expected_turn_on_equation}, given a $\phi_{off}$, 
length of the null (NL), P$_3$,  and slope of driftband (D) before the null.
\begin{equation}
 \phi_{on-exp} ~  = ~ \phi_{off} + mod(D\times{NL}, P_3)  
 \label{expected_turn_on_equation}
\end{equation}
Here, $mod$ represents remainder from the drifting periodicity. If we find the $\phi_{on-exp}$
to be outside the width of the component, we subtract the width of the corresponding component to 
truncate $\phi_{on-exp}$ inside the component width window. In principle, the bounding limits for sub-pulse 
localization should come from measurement of P$_2$. However, for example, the leading component of \psrb\ 
has $\hat{D}$ of around 0.9${^\circ}\pm$0.4$^\circ/P_1$
and P$_3$ of around 13.5$\pm$0.7 periods, we can find that P$_2$ would be 12${^\circ}\pm$6$^\circ$ 
which is equivalent to its width (11${^\circ}\pm$0.5$^\circ$) with larger errors. 
Thus, we chose to use component widths as more suitable bounding limits here in our case.
We calculated $\phi_{on-exp}$ for a range of null lengths for both pulsars. 

For \psra, the absolute phase differences between the expected and observed turn--on 
phase ($\bigtriangleup{\phi}~=~|\phi_{on-exp}~-~\phi_{on}|$) is shown in 
Figure \ref{stopage_drifting}. If drifting continues at its regular speed with a memory of turn--off phase, 
one is likely to observe roughly zero $\bigtriangleup{\phi}$ for a given null length.
From Figure \ref{stopage_drifting} it can be pointed out that drifting does not continue 
at its regular speed during the observed null lengths. 
If \psra\ exhibits phase memory only for shorter nulls, we can expect to find 
smaller $\bigtriangleup{\phi}$ for shorter nulls. However, $\bigtriangleup{\phi}$ 
are almost equal to the width of the component across various null lengths without exhibiting any {\br trends} 
from shorter to longer nulls. Thus, it can be speculated that \psra\ does not retain information regarding the 
turn--off phase or drifting slopes of burst pulses prior to nulls. 
The turn--on phases are also random and do not have any association with the 
turn--off phases as pointed out in the earlier section. Thus, we can speculate that \psra\ either slows down 
or speeds up its sub--pulse drifting during the null state. However, it should be noted that the duration 
of our observations and sensitivity of single pulses were not sufficient 
(as is clear by the relative error bars in Figure \ref{turnoff_turnon}) 
to investigate this in detail. Higher sensitive and longer observations are 
required to investigate changes in the drift rate and its association with 
larger number of null transitions in this pulsar. 

\begin{figure}
 \centering
 \subfigure[]{
 \includegraphics[width=3 cm,height=8 cm,angle=0,bb=0 0 425 999]{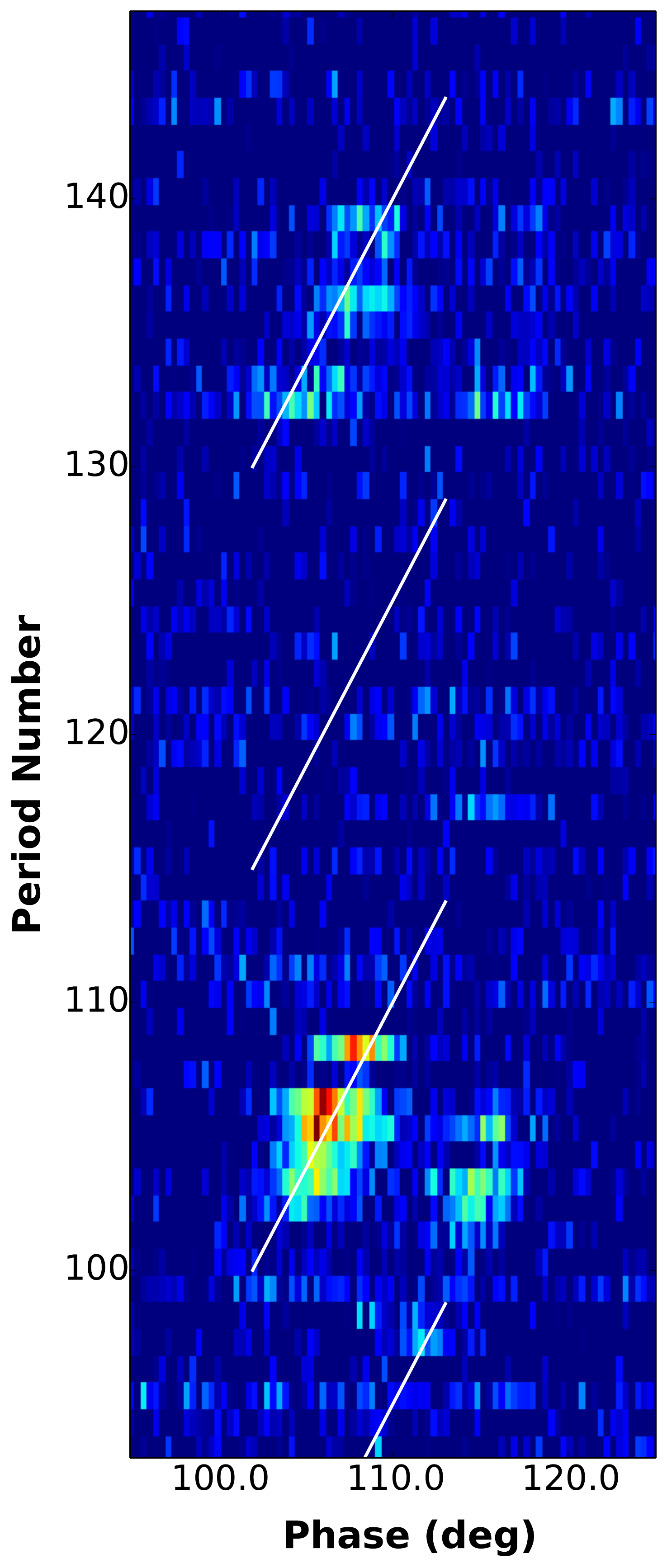}
 \label{simulation_fit_fig1}
 }
 \centering
 \subfigure[]{
 \includegraphics[width=3 cm,height=8 cm,angle=0,bb=0 0 425 1009]{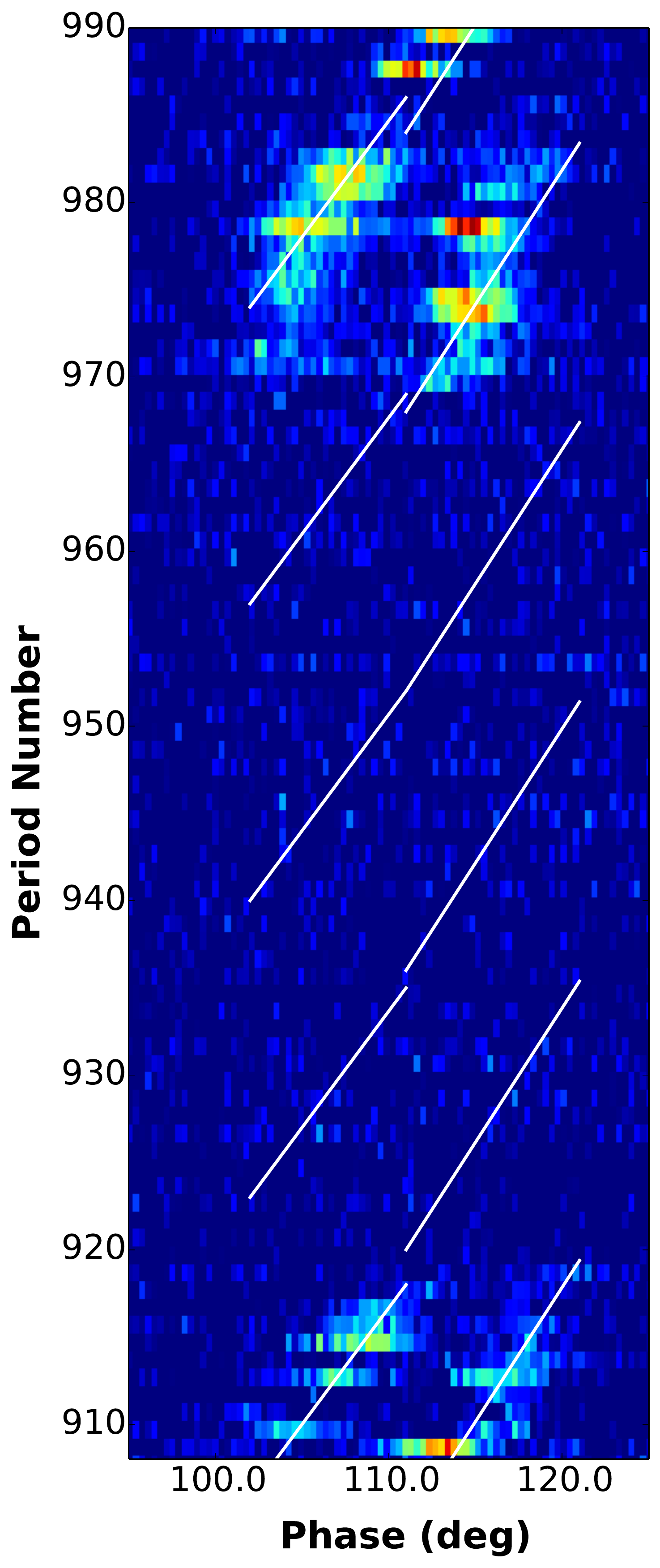}
 \label{simulation_fit_fig2}
 }
 \caption{Examples showing speculative drifting during nulls for the leading and trailing components of \psrb\ at 625 MHz with 
 (a) one missing driftbands (only shown for the leading component) and (b) three missing driftbands. 
 The straight lines here are not fit to the driftbands in this 
 case but are shown here just to illustrate approximate locations of the missing driftbands.}
\end{figure}

Contrary to \psra, when similar $\bigtriangleup{\phi}$ were derived for the leading component of \psrb, 
we found remarkable similarities between the predicted and the measured turn--on phases. 
Figure \ref{Stopage_drifting_1840} shows these $\bigtriangleup{\phi}$ 
for various null lengths. We found that the $\bigtriangleup{\phi}$ are significantly 
smaller than the width of the component and are located close to zero phase difference within 
the errors. Similarly, for the trailing component of \psrb, we also found that the distribution of 
$\bigtriangleup{\phi}$ is around zero phase difference (see Figure \ref{Stopage_drifting_1840_trail}). 
As noted earlier, the trailing component is relatively weak and sometimes does not provide good fits to the 
drift--rate slopes, thus we were able to measure the $\bigtriangleup{\phi}$ for only eight transitions. 
These transitions were all came from mid to longer null lengths as shown in Figure \ref{Stopage_drifting_1840_trail}.
It should be noted that as D$_l$ and D$_t$ exhibit large spread in the measurements, 
we consider D$_l$ and D$_t$ before the onset of nulls and assume that they are constant across null states for 
the above calculations for the leading and trailing components, respectively. 
These assumptions may not be accurate for a few null states, which 
could explain the small spread from zero phase difference we see in Figures \ref{Stopage_drifting_1840} and \ref{Stopage_drifting_1840_trail}. 
Considering these small differences, these results still suggest that \psrb\ is likely to be among those pulsars {\br that exhibit continuous drift in} 
sub--pulses during nulls as it {\br exhibits} remarkable memory of its emission phase across the nulls in both 
of its profile components. Moreover, as pointed out in 
an earlier section, the turn--off phases are most likely to occur near the trailing-edge of 
component (end of driftband for positively drifting sub--pulses) and turn--on phases are likely to occur near the 
leading-edge of the component (start of the driftband for positively drifting sub--pulses). This leads to a 
speculation that, {\itshape is it likely that nulls in \psrb\ represent a complete absence of an entire driftband? }

Figures \ref{simulation_fit_fig1} and \ref{simulation_fit_fig2} 
show examples of two such null instances with missing one and three driftbands, respectively. 
An example of weak trailing component post short null can be seen in Figure \ref{simulation_fit_fig1}
where making a true judgment regarding the turn--on phase was not possible.
It should be noted that these examples shown here are just illustrations for possible locations 
of missing driftbands. The straight lines shown do not represent real fit to these driftbands. 
An additional confirmation regarding the missing driftbands can be considered from the 
measured null--length distributions shown in Figure \ref{NL_BL_1840}. 
This distribution clearly shows an excess of around 12 to 18 period nulls which is equivalent to 
the drifting periodicities of 13.5$\pm$0.7 periods for the leading component and 
18$\pm$1 period for the trailing component.       
This further supports our findings that most of the nulls in the leading and trailing components 
of \psrb\ are missing driftbands. However, these P$_3$ measurements, as noted 
in Section \ref{drifting_sect}, could easily be influenced by intervening nulls 
and thus {\br may not} be accurate. As the pulsar has long period and our sensitivity 
was limited we were only able to study this behavior for a handful of driftbands in \psrb. 
Higher sensitivity and long observations are required to scrutinize 
it further for both components.

\section{Discussion and conclusions}
\label{discussion_and_conclusion_sect}
We have reported nulling, drifting and their interactions in PSRs \pa, and \pb. 
Both pulsars clearly exhibit nulling and prominent drifting behaviors.   
A brief summary of our findings {\br is} listed in Table \ref{Table_of_results}. 
For \psra\ we estimate NF to be around 30$\pm$5\% while \psrb\ exhibits large nulling 
with NF of around 50$\pm$6\%. In a few nulling pulsars, it has been shown that the previously 
reported nulls are not true nulls. For example, PSR J1752+2359 appears to show weak single 
burst pulses during the null states which have significantly different polarization properties compared to the average profile 
\cite[]{gjw14a}. In order to scrutinize any weak level emission during the observed nulls, 
we separated all the nulls and burst pulses for both pulsars. We did not find any detectable emission after combining 
all null pulses together for either pulsar. Thus, we can conclude that these nulls are true 
nulls and are not weak emission states (within the sensitivity limit of the observations).

\subsection{Are different P$_3$ modes real?}
We found that the trailing component of \psra\ clearly exhibits driftbands while the leading 
component exhibits amplitude modulation without clearly exhibiting any drifting sub--pulses. 
We measured this amplitude modulation to be around 4.9$\pm$0.1 periods for the leading component while 
the trailing component showed two distinct periodicities of around 4.3$\pm$0.1 periods and 4.6$\pm$0.2 periods. 
These measurements match with the previously reported periodicities for this pulsar \cite[]{wes06,wse07,bmm16}. 
In order to verify the different P$_3$ modes, we measured periodicities for each consecutive and overlapping blocks of 356 pulses. 
We did not find any blocks that display dominating periodicity mode implying that these changes could occur 
on a smaller time-scale than the size of the block we have selected here. In order to scrutinize these changes, 
we measured slopes of driftbands which showed a very large spread around the mean value of 1.1$^{\circ}\pm$0.2$^\circ/P_1$. 
This spread was large enough to generate different P$_3$ periodicities seen in \psra. Thus, we conclude that drifting in \psra\ 
is very sporadic and it changes irregularly without clearly producing distinguishable drift modes. 

Drifting in \psrb\ is very intriguing with relatively broad periodicity of 13.5$\pm$0.7 periods in the leading component. 
The trailing component showed a broad P$_3$ that peaks around 18$\pm$1 periods. We noticed that these periodicities in either 
components are not stable and show large spread in the measured drift slopes. Although the distribution of 
D appears to be broad, we carried out {\itshape Hartigans' Dip test for Unimodality} and found that the distribution is 
still unimodal with high significance. In order to avoid small sample bias, we generated a range of slopes using Monte-Carlo 
simulation but did not find strong case for bi--modality. We measured the average slope to be around 0.9$^{\circ}\pm$0.4$^\circ/P_1$ 
and 0.7$^{\circ}\pm$0.4$^\circ/P_1$ for the leading and trailing components in \psrb, respectively.
We can clearly see that the driftband slopes are relatively similar for both components in \psrb\ and the apparent 
difference seen in P$_3$, measured from LRFs, are likely originated from intervening nulls and sporadic nature of 
the driftband slopes. For example, the errors on D$_l$ and D$_t$ can easily produce this different P$_3$ values when 
measured for a large number of driftbands. We also measured slopes of driftbands exclusively around the nulls for both pulsars but did not 
find their distributions to be significantly different from the general distributions for either pulsars. Thus, we conclude that 
PSRs \pa\ and \pb\ do not exhibit noticeable drift slope changes around nulls. 

These results highlight an important point which should be noted that P$_3$ is an average quantity which requires several 
pulses that follow similar repeating pattern. However, pulsars like PSRs \pa\ and \pb\ indicate that changes in the 
drifting patten could occur on a much smaller time-scale (less than the size of a single driftband). 
Thus, measuring the slope of a driftband is essential to scrutinize any changes in the drifting pattern. 
Such study needs to be conducted for more drifting pulsars which exhibit different P$_3$  
and/or P$_3$ modes between different profile components. In other words, the apparent P$_3$ differences
could simply be due to rapidly changing driftband slopes.  
Furthermore, \cite{bmm16} have also suggested that in a few instances P$_3$ measurements could cause ambiguity 
due to aliasing, which eventually hinders our understanding in connecting the observed behavior with the actual physical phenomena. 
Recently, \cite{mbt17} also indicated that different P$_3$ modes seen in PSR B0031$-$07 were actually 
gradual changes in the driftband slopes. Thus, we would like to emphasize that the different P$_3$  
needs to be scrutinized carefully by measuring changes in their driftband slopes as P$_3$ could lead to 
incorrect interpretation of the physical changes. 
 
\begin{table}
\centering
\caption{A summary of results}
 \begin{threeparttable}
 \begin{tabular}{l c c}
 
 \hline
 PSR 		 					& \pa\       	& \pb \\
 \hline
 \hline
 Nulling Fraction (\%)	 				& 30$\pm$5  	&  50$\pm$6     \\
 Quasi-periodicity (periods)   				& 33$\pm$4  	&  $-$          \\
 Leading comp. P$_3$ (periods)				& 4.9$\pm$0.1   &  13.5$\pm$0.7 \\   
\multirow{2}{*}{Trailing comp. P$_3$ (periods)}     	& 4.3$\pm$0.1   &  \multirow{2}{*}{18$\pm$1}  \\
							& 4.6$\pm$0.2   &  		\\
Leading comp. $\hat{D}$ ($^\circ/P_1$) 			&  --		&  0.9$^{\circ}\pm$0.4$^{\circ}$ \\
Trailing comp. $\hat{D}$ ($^\circ/P_1$)  ...		& 1.1$^{\circ}\pm$0.2$^{\circ}$   &  0.7$^{\circ}\pm$0.4$^{\circ}$  \\
$\phi_{on}$  leading comp.				& $-$		&  103.9$\pm$0.2 \\
$\phi_{off}$ leading comp.				& $-$		&  106.9$\pm$0.1 \\		
$\phi_{on}$  trailing comp.				& $-$		&  113.2$\pm$0.3 \\
$\phi_{off}$ trailing comp.				& $-$		&  117.7$\pm$0.2 \\							
 $\bigtriangleup{\phi}$	 				& $\sim{W_c}$   &  $\ll{W_c}$    \\
 Phase memory across nulls  				& No		&  Yes		 \\
 \hline
\end{tabular}
\label{Table_of_results}
\begin{tablenotes}[para,flushleft]
 \small
 \item Note : The $\hat{D}$ represents average slopes while the $\bigtriangleup{\phi}$ represents absolute phase difference 
 between $\phi_{on}$ and $\phi_{on-exp}$ in term of width of the corresponding component ($W_c$). 
\end{tablenotes}
\end{threeparttable}
\end{table}

\subsection{On the origin of pulsar nulls}
Among the various observed phenomena of radio pulsars, pulse nulling remains 
one of the most intriguing ones as no reasonable model has been able to explain the wide variety of observational 
features. Models such as temperature fluctuations to change coherence 
conditions \cite[]{che81,dchr86}, switching between different gap discharge mechanisms \cite[]{dh86,zqlh97,zqh97b}
and changes in the magnetospheric currents \cite[]{tim10} link pulsar nulling with the intrinsic 
magnetospheric changes. There are also other models which connect pulsar nulling with geometry of the pulsar beam \cite[]{hr07,hr09,rw08}. 
The emission in the radio beam, as suggested by the \cite{rs75}, comes from 
localized regions (\emph{aka.} sparks) on the polar cap. This notation was further 
investigated and a comprehensive model for subpulse drifting was suggested as a rotating 
\emph{carousel} of sub-beams \cite[]{ran93,dr99,dr01}. By treating the sub-beams as organized in an evenly spaced 
uniform pattern, \cite{hr07} argued that empty line--of--sight cuts or extinguished individual sub--beams 
are likely to repeat after certain pulsar periods thus giving rise to periodic nulling. 

For \psra, we found a clear quasi--periodic fluctuation of around 33$\pm$4 periods in the 
burst bunching. Such quasi--periodic pattern in the occurrence of nulls have been reported in a few pulsars 
such as PSRs B0834+06 \cite[]{wr07}, B1133+16 \cite[]{hr07}, J1738$-$2330, and J1752+2359 \cite[]{gjw14a}. 
This quasi-periodic bunching in \psra\ can also be confirmed from its distributions of 
clustering of null--length and burst--length. The empty line--of--sight model also suggests 
that, as the sub-beams continue to drift during nulls, one would expect to predict the 
turn--on phase with high accuracy. In order to scrutinize drifting during nulls, we 
measured expected turn--on phase for nulls with various lengths for both pulsars. 
We found that \psra\ does not seem to have any memory of the turn--off phase during 
the nulls as the absolute differences are comparable to the width of the component 
and did not show any {\br trends} for longer or shorter nulls. Furthermore, missing 
line-of-sight model can explain certain fraction of partial nulls $-$ nulls seen in only 
one of the profile components \cite[]{la83,viv95,vkr+02,jv04}. 
In the case of \psra, we did not notice any partial nulls. 
Moreover, recent studies by \cite{gjk12} and \cite{cor13} 
showed that nulling is inherently random and state transition from null-to-burst, and vice-verse, can be 
modeled by Markov chain models. \cite{cor13} further suggested that certain external influences  
acting with these random emission state transitions could mimic quasi-periodic energy fluctuations. 
Such influences may arise from an external body \cite[]{csh08} or neutron star oscillations \cite[]{cr04a} 
or near-chaotic switches in the magnetosphere's non-linear system \cite[]{tim10}. 
Thus, base on lack of phase memory and partial nulls, we speculate that the missing 
line-of-sight model is less likely to produce the quasi-periodic feature seen in \psra. 

We also studied preferences for turn--on and turn--off phases around nulls for both pulsars. 
For \psra\ we did not find any peculiar preference for these phases. We also measured 
cross-correlation between the measured phases but did not notice any {\br trends} which might suggest 
a complete stop in drifting during nulls and resume from the same turn--off phase after nulls. 
However, for \psrb\ we found strong preferences for turn--on and turn--off phases which are significantly 
different from each other. It seems that, in most occasions \psrb\ tends to start nulling after what appears to be an end of a driftband for both components.
Thus, most of the turn--off phases were found to occur near the trailing edge of a given component. 
Similarly, when the pulsar switches back to emission phase, in most occasions it starts at the beginning of a new driftband 
as most of the turn--on phases were {\br found near} the leading edge of both components. 
Such preferences have not been seen in any nulling-drifting pulsar to the best of our knowledge. 
Moreover, \psrb\ also showed a good match between the expected and measured turn--on phases which 
seems to suggest that it might continue to drift during nulls. Furthermore, \psrb\ also shows a good 
fraction of nulls with length similar to the length of a typical driftband. 
Thus, it is likely that nulling in \psrb\ is due to extinguishing sub-beams 
and empty line-of-sight passing over these sub-beams as proposed by \cite{hr07}.
If the above model is true for \psrb, one would also expect to see partial nulls as both components 
are likely to be originating from the outer conal ring. However, both components seem to show simultaneous 
nulling as seen from the single pulses in Figure \ref{1840_spdisplay}. 
However, we can not completely rule out missing line-of-sight model in this case due to lack 
of our knowledge in the localization of these components in the emission cone.

Assuming that \psrb\ does retain information regarding the phase of the last burst pulse, 
an alternative theory to missing line--of--sight model can be explored. This model of pulse nulling, 
first proposed by \cite{fr82}, predicts a break
in the two--stream instabilities which occurs during a steady polar gap discharge. 
Two--stream instabilities are produced due to the propagation of secondary particles with different 
momenta. These secondary particles are generated from the high--energy primary 
particles at initial gap discharge. A situation can occur
for long period pulsars (like \psrb) in which gap discharges, normally at roughly the same rate as the potential drop,
would increase before the sparking. The polar cap then attains a steady discharge state in which 
no high-energy primary particles are produced for subsequent consecutive two--stream instabilities and bunching. 
It can be speculated that, this steady state is reached after a buildup from the residual potential, 
which is a {\itshape left-over} from each gap discharge during normal non-steady condition (non-null state). 
Once the residual potential reaches the maximum gap height potential, the pulsar attains a steady state
during which no radio emission is emitted (null state). However, the sparks could still be rotating on top of the polar cap. 
When the residual potential discharges, the pulsar resumes its normal interrupted sparking action 
that gives rise to primary particles with high energy for the consecutive bunching and coherence. The charging and discharging 
of this residual potential, in addition to the normal gap discharge, has an inherent 
memory associated with them. The time-scale of this memory mechanism should be higher 
for pulsars with long period and low surface magnetic field. \psrb\ here fits the criteria (see Table \ref{intro_table})
for this mechanism to be operational which could be the reason we are seeing strong 
evidence of such phase memory during its nulls. 
 
Although, we are able to demonstrate remarkable nulling and drifting interactions in \psrb, more 
sensitive and longer observations would certainly provide intriguing details. Recently, the GMRT has been upgraded 
with the wider bandwidth of 200 MHz \cite[]{kum14,kbb16} which is much broader than 32 MHz used for our observations.
Observations using such wider bandwidth will improve single pulse S/N by orders of magnitude and 
will allow us to robustly model both the components. 

\section{Acknowledgments}
We would like to thank the referee for very useful comments and suggestions. 
VG acknowledge the West Light Foundation
of the Chinese Academy of Sciences project XBBS-2014-21.
VG would also like to thank Bhal Chandra Joshi and Dipanjan Mitra for useful discussion 
regarding pulsar drifting in general. JPY supported by the Strategic Priority Research 
Program of the Chinese Academy of Sciences, Grant No. XD23010200. RY acknowledges supports 
from NSFC Project no. 11573059; the Technology Foundation for Selected Overseas Chinese Scholar, 
Ministry of Personnel of China; and the Joint Funds of the NSFC Grant No. U1531137. 
We thank the staff of the GMRT who made these observations
possible. The GMRT is operated by the National Centre for 
Radio Astrophysics of the Tata Institute of Fundamental Research
(NCRA-TIFR).

\bibliographystyle{aasjournal}
\bibliography{modpsrrefs,mybib,modrefs,psrrefs}

\end{document}